\documentclass[preprint]{imsart}

\RequirePackage{amsthm,amsmath,amsfonts,amssymb}
\usepackage{amsmath,amssymb}
\usepackage{stackengine,graphicx}
\usepackage[dvipsnames]{xcolor}
\usepackage{bbm}

\usepackage{algorithm}
\usepackage{algpseudocode}

\usepackage{collectbox}
\usepackage{bbm}
\usepackage{capt-of}

\usepackage{hyperref}
\hypersetup{
    colorlinks=true,
    linkcolor=blue,
    filecolor=magenta,      
    urlcolor=cyan,
    citecolor=black
}

\usepackage[framemethod=TikZ]{mdframed} 
\usepackage{tcolorbox} 

\def\l{\left}
\def\r{\right}

\newcommand{\f}{\frac}

\usepackage{array}
\newcolumntype{L}{>{\arraybackslash}m{7cm}}

\def\adaptNon{\textbf{adaptive$_{\text{non}}$ }}
\def\adaptOpt{\textbf{adaptive$_{\text{opt}}$ }}
\def\adaptOver{\textbf{adaptive$_{\text{over}}$ }}

\usepackage[margin=0.5in]{geometry}

\begin{document}

\begin{frontmatter}

  \title{Adaptively stacking ensembles for influenza forecasting with incomplete data}

  \begin{aug}
    \author[A]{\fnms{Thomas} \snm{McAndrew}\ead[label=e1]{mcandrew@umass.edu}},
    \author[A]{\fnms{Nick} \snm{Reich}\ead[label=e2]{nick@schoolph.umass.edu}},

    \address[A]{ Department of Biostatistics and Epidemiology,\\
School of Public Health and Health Sciences\\
University of Massachusetts~Amherst,\\
Amherst, Massachusetts, United States of America \printead{e1}}

  \end{aug}
    
\begin{abstract}
Seasonal influenza infects between $10$ and $50$ million people in the United States every year. 
Accurate forecasts of influenza and influenza-like illness (ILI) have been named by the CDC as an important tool to fight the damaging effects of these epidemics. Multi-model ensemble forecasts have shown positive results.
But ensemble forecasts of influenza outbreaks have been static, training on past ILI data, generating a set of optimal component model weights, and keeping the weights constant.
We propose an adaptive ensemble that (i) updates weights week-by-week throughout the influenza season, (ii) only needs the current season's data to make predictions, and (iii) by introducing a prior distribution, shrinks weights toward an equal weighting.
We investigate the prior's ability to impact adaptive ensemble performance and, finding an optimal prior, compare our adaptive ensemble’s performance to equal-weighted and static ensembles.
Applied to forecasts of short-term ILI incidence at the regional and national level, our adaptive model outperforms an equal-weighted ensemble and has similar performance to the static ensemble, needing years of training data.
Adaptive ensembles can quickly train and forecast during epidemics, providing a practical tool to public health officials looking for forecasts that can conform to unique features of a specific season. 
\end{abstract}


\end{frontmatter}

\section{Introduction}

Every year seasonal influenza costs hospitals time and resources, and causes significant loss of life, especially among patients with cardiac disease, previous respiratory illness or allergies, children and the elderly~\cite{centers2009prevention,russell2008global,harper2009seasonal,garten2018update}.
During the season's peak hospitals admit patients beyond capacity, and in severe cases, for greater lengths of stay, postponing lower priority but still important surgeries~\cite{krumkamp2011health,schanzer2013impact,molinari2007annual}.
The total economic burden of influenza outbreaks (including medical costs, lost earnings, and loss of life) in the US alone are estimated to exceed \$85 billion annually~\cite{molinari2007annual}.

Accurate forecasts of influenza can help with public health response.
Public health officials at the national, regional, and state level benefit from the advance warning forecasts can provide, by preparing hospitals for increased disease burden~\cite{reed2009infection}, shifting vaccines to places in need~\cite{germann2006mitigation}, and alerting the public~\cite{vaughan2009effective}.
Recognizing the value of accurate forecasting, the US Centers for Disease Control and Prevention (CDC) have organized forecasting challenges around seasonal influenza outbreaks~\cite{biggerstaff2016results,biggerstaff2018results,mcgowan2019collaborative}. 
These challenges focus on generating weekly forecasts of influenza-like illness (ILI) rates at the national and regional level in the US. 
ILI is a syndromic classification and defined as a fever plus an additional symptom such as a cough or sore throat.
Public health officials regard ILI surveillance data as an important indicator of the trends and severity of a given influenza season~\cite{biggerstaff2017systematic}.

The CDC influenza forecasting challenges, called FluSight, have spurred methodological development in the area of influenza forecasting and infectious disease forecasting more broadly.
This effort has led to multiple research groups developing models trained on historical ILI data to forecast future incidence across the US.
These models include statistical and mechanistic models, among other approaches~\cite{shaman2012forecasting,brooks2015flexible,ray2017infectious,osthus2018dynamic}.

Decision-makers have requested a single unified forecast, and several groups have developed multi-model ensemble forecasts of influenza. 
These models synthesize predictions from a set of different component models and have shown better predictive ability compared to any single model~\cite{yamana2017individual,ray2018prediction,reich2019collaborative}.

Multi-model ensembles aggregate diverse component model predictions into a final `meta-prediction' that is often, but not always, more robust than any single component model~\cite{zhou2012ensemble,sewell2008ensemble}.
This robustness comes from the ensemble's mixture of component models.
Freed from having to make a single set of assumptions about the data, the ensemble never depends on just one model and associated limitations~\cite{zhou2012ensemble,sewell2008ensemble}.
In contrast to ensemble modeling approaches like 
bagging~\cite{breiman1996bagging,dietterich2000ensemble} and boosting~\cite{schapire2003boosting,dietterich2000ensemble} which create a library of similar component models, multi-model ensembles combine distinct component models together, possibly models designed and implemented in completely different modeling paradigms. Application examples include weather \cite{krishnamurti1999improved}, economics\cite{garratt2011real}, energy production\cite{pierro2016multi}, education\cite{adejo2018predicting}, and infectious disease \cite{reich2019collaborative}. 

The multi-model ensembles addressed in this work are related, but conceptually distinct from Bayesian Model Averaging~(BMA)~\cite{raftery2005using,steel2011bayesian,madigan1996bayesian,montgomery2010bayesian}.
BMA places a prior over models that are proposed to have generated the observed data~\cite{raftery2005using}.
BMA assumes, that from a set of possible models, a single model generated the observed data, and the goal of BMA is to collect data and draw inference about which single model generated the entire data stream.
The multi-model ensembles in this work take a different approach~\cite{minka2000bayesian,bishop2006pattern}.
These models assume that every observation was generated by a probability distribution whose true form is a combination of multiple distributions.
In an ensemble like this, the goal is to determine how the models are combined to from a single data generating distribution---no single model is assumed to generate the entire data stream.

Extending an existing ensemble implementation~\cite{reich2019collaborative}, we developed a new method for combining component models that relies on recently observed in-season data to adaptively estimate a convex combination of models, sometimes referred to as a ``linear pool''.
Though past work has studied methods for training ensembles from a fixed data set~\cite{gneiting2005weather,foley2012current,van2007super}, and continuously adapting ensembles to streaming data~\cite{pari2018multi,fern2003online,benkeser2018online},
a unique and new challenge in our setting is the reliance on noisy, revision-prone observations like weekly ILI surveillance data.
Recent ILI observations, because they come from a real-time public health surveillance system, are subject to revisions, which are occasionally substantial and can negatively impact forecast accuracy ~\cite{osthus2019even,reich2019collaborativepnas}.
Like past work in ensemble forecasting, this work combines a fixed set of probabilistic forecasting models provided by different forecasting teams~\cite{reich2019collaborativepnas} and does not presume to have the capability to refit  component model parameters.

To protect the adaptive multi-model ensemble framework from relying too heavily on recent, revision-prone ILI data, we developed a Bayesian model combination algorithm that uses a prior to regularize ensemble weights.
Previous results in this field show equally-weighted ensembles often perform very well~\cite{mcgowan2019collaborative,reich2019collaborative}, and to align with these results, we chose a uniform Dirichlet prior that shrinks model weights towards equal.
Our prior is also time dependent. 
Unlike a typical model where the prior becomes less influential with increasing data, our prior is designed to exert a constant influence on the final model weights. 
Our model never allows model weights to depend only on revision-prone data.
We show our Bayesian framework can be solved with a Variational Inference algorithm~(\textbf{devi-MM}), and demonstrate that an existing Expectation-Maximization approach to ensemble weight estimation~\cite{reich2019collaborative} is a special case.

We compared ensembles that assign equal weights to models~(\textbf{EW}) and those with possibly unequal but static and unchanging weights throughout a season (\textbf{static}),  against our adaptive ensemble~(\textbf{adaptive}) which updated component model weights every week.
Static ensembles were trained on all cumulative forecasting data before the beginning of each season and were unable to modify their model weights in-season.
In contrast, adaptive ensembles were only trained on within-season data but could modify their weights as each additional week of ILI data was reported.
Comparing static and equal to adaptive ensembles, we highlight: (i) the adaptive model's ability to buffer against sparse and revision-prone data by using a prior over ensemble weights and (ii) similar, and sometimes improved, performance when comparing adaptive to static models, where adaptive models require substantially less data to train.

The manuscript is structured as follows: Section~2 describes the CDC influenza training data and component models included in all ensembles, defines performance metrics, and develops the methodology for our adaptive ensemble. Section 3 presents results, investigating the prior's impact on adaptive ensemble performance and compares the adaptive ensemble to static and equally weighted ensembles. Section 4 relates this work to linear pooling algorithms, and discusses our approach in the broader context of data-driven, evidence-based public health decision making.

\section{Methods}

\subsection{Data}
\subsubsection{Influenza data}
Every week throughout the season, for $10$ different reporting regions (HHS1, HHS2, ... , HHS10) and a national average, the CDC publishes surveillance data on the number of patients who visited a sentinel site and the number of patients that were diagnosed with influenza-like illness~(ILI).
Percent ILI is defined as the number of patients presenting with a fever~(greater than 100F) plus cough or sore throat divided by the number of all patient visits times one hundred.
More than $3,500$ outpatient clinics report ILI percentages to the CDC as part of the ILINet surveillance system.
In addition to the newest ILI data, updates to past ILI percentages are also published weekly.

Because of reporting delays, ILI percentages undergo revisions every week after they're first reported, finalized only several weeks after the season ends.
Revisions make predicting future ILI difficult. Component model forecasts need to update week-to-week for two reasons: (i) the newest ILI data is reported, and (ii) all previous ILI values have been revised.
When forecasting models attempt to adjust for this revision process and the uncertainty in reported data they typically outperform models that do not account for data revisions~\cite{brooks2018nonmechanistic}.

\subsubsection{Forecasting data}

The FluSight Network (FSN) is a collaborative group of influenza forecasters that use historical ILI data to build retrospective forecasts and the performance of models on this retrospective data to build ensemble forecasts.
Teams train their models and generate forecasts as if their model had been applied in real-time across all seasons since $2010/2011$.
The resulting collection of forecasts serves as a historical record of how each model would have performed in real-time~\cite{reich2019collaborativepnas}.

Probabilistic forecasts for $21$ `component' models were submitted to the FSN for the $2018/2019$ season.
The majority of forecasts were submitted retrospectively (models from teams KoT and CU submitted in real-time), as if they were forecasting in real-time and contending with ILI data revisions.

\begin{figure}[ht!]
  \centering
   \includegraphics[scale=0.60]{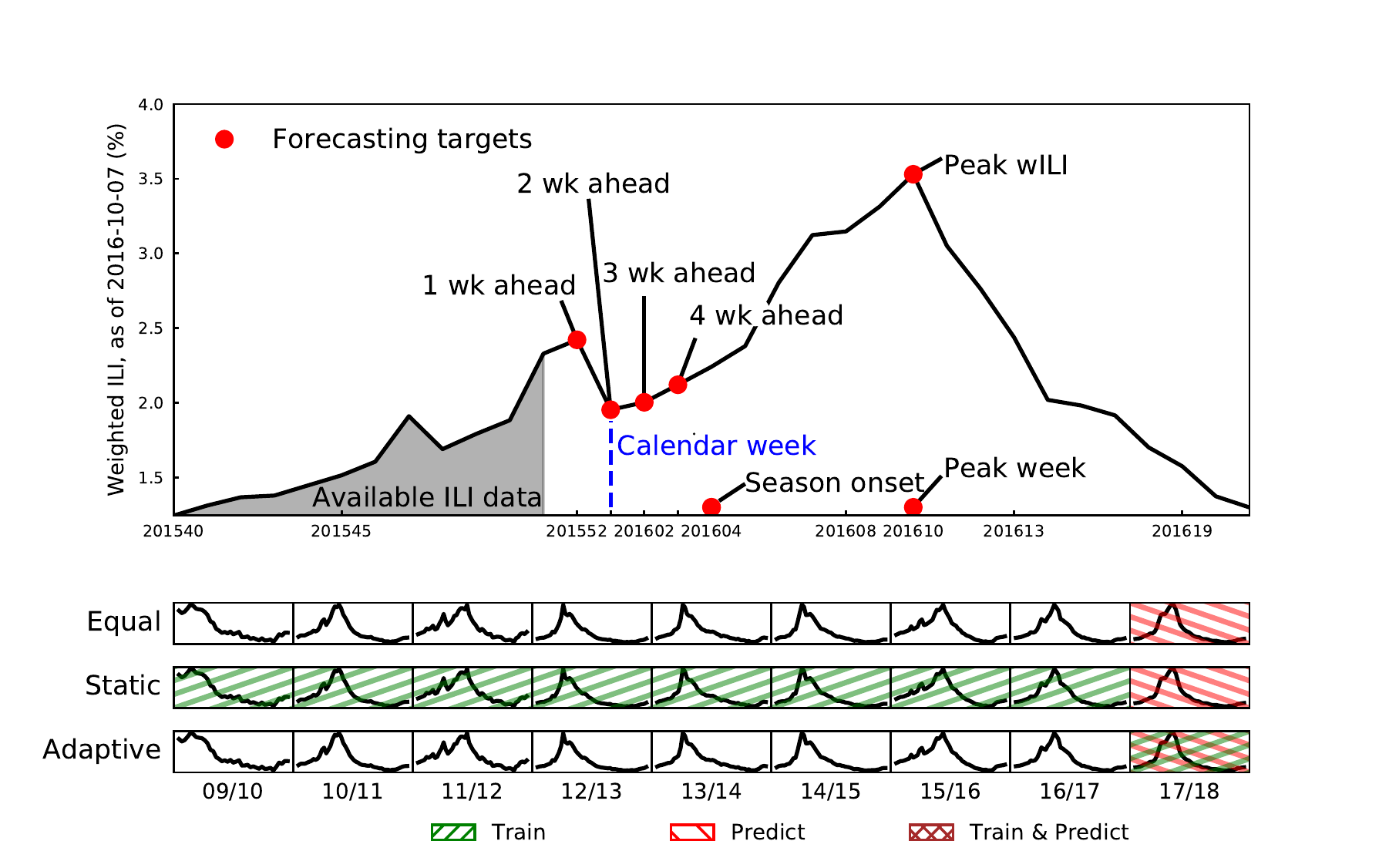}
  \caption{
 Overview of forecasting targets and data used in model estimation. {\bf Top Panel:} Zooming in on one region-season illustrates the $7$ forecasting targets, divided into week ahead ($1$, $2$, $3$, and $4$), and seasonal (onset, peak week, peak percentage) targets.
    Due to delays in reporting at time $t$, forecasting models only have data available up to $t-2$ weeks, making the current calendar week the $2$~wk ahead forecast (sometimes titled the nowcast). 
    {\bf Bottom Panels:} Three ensemble algorithms: equally-weighted, static, and adaptive, take different approaches to training and prediction, illustrated by how they would operate in the 2017/2018 season. 
    Equally-weighted ensembles ignore past ILI data, weighting all component models the same and forecast the next season.
    Static ensembles train on all previous seasons~(green, up-right slanting lines), find optimal weights for all component models, and keeping these weights fixed, forecast the next season~(red, down-right slanting lines).
    Adaptive ensembles train and forecast all in the current season~(red and green cross-hatch), ignoring any past data.
    For adaptive ensembles, every week component model weights are updated on all current within-season data and then used to forecast the next week.
    \label{fig1.forecastingConcept}}
\end{figure}

For the CDC FluSight challenges, component forecast models and ensembles built from them, forecast $7$ key targets (Figure \ref{fig1.forecastingConcept}). These targets are $1$, $2$, $3$, $4$ week-ahead ILI percentages, onset week (defined as the week where $3$ consecutive weeks are above the CDC-specified baseline ILI percentage), the season peak week (the epidemic week with highest ILI percentage), and the season peak percentage (the highest ILI percentage throughout the season).
The CDC cannot determine the start of the influenza season and when the peak occurs until mid-summer due to continued revisions to data.
This means final performance data on forecasts---the data used for building ensembles---for the $3$ seasonal targets~(baseline onset, season peak week and percentage) can not be compiled until after the season closes.
Because our adaptive ensemble is trained on within-season data, only component model forecasts of the $4$ week-ahead targets will be used for training from the FluSight Network.

The difficulty estimating ensemble weights comes from ILI data revisions.
An adaptive ensemble must compute weights based on component model performance---the probability component model forecasts place on true ILI values.
But past ILI data is revised weekly and changes component model performance in past epidemic weeks.
Unlike an ensemble that trains on component models scored on finalized ILI data, an adaptive ensemble must account for ILI revisions because they change past component model performance week to week.

Seasons from $2011/2012$ up to $2017/2018$ will be used to compare equal, static, and adaptive ensembles. The $2010/2011$ season will be used as hold-out data to build a prior for the adaptive ensemble.

\subsubsection{Performance metrics} \label{performanceMetrics}
A useful predictive model must be well-calibrated and accurate, consistently placing high probability on potential ILI percentages, that after all revisions, become the final ILI percentage.
A proper scoring rule~\cite{dawid2014theory} widely accepted in the forecasting community is the logscore~\cite{dawid2014theory,reich2019collaborativepnas}, and defined as the predicted probability placed on a future true (or finally revised) ILI value.
The CDC adopted a modified logscore as their standard scoring metric.
Given a model with probability density function $f(z)$ and true ILI value $i*$, both the logscore and modified logscore can be written
\begin{equation}
  \text{logscore}\l(f\r) = \log \l( \int_{z = i*-w}^{z=i*+w}f(z)\; dz \r), \label{analyLogScore}
\end{equation}
where $w$ is a small positive number.
The traditional logscore sets $w=0\%$ and this is the metric we use to compare ensembles.

In practice, we bin predictive probabilities of ILI percentage from $0$ to $13\%$ by $0.1\%$ plus a single bin from $13\%$ to $100\%$.
The proper logscore reduces to computing the log of the probability given to the single true ILI bin
Models that place high probabilities on the bins containing the observed final ILIs are scored closer to $0$ (perfect score), while models that place low probabilities on observed ILI bins are scored closer to $-\infty$ (the worst score).
Logscores are truncated to -10 to follow convention from CDC FluSight scoring~\cite{mcgowan2019collaborative}.

\subsection{Ensemble Framework}
\subsubsection{Ensemble model specification}
Static and adaptive ensemble algorithms use a mixture model to combine (or average) component models forecasts. 
We developed two approaches to finding ensemble weights: a maximum likelihood method and a maximum aposteriori Bayesian approach. 
A Bayesian approach lets us regularize component model weights by modifying our log likelihood with a prior.
The choice of prior for our adaptive ensemble is meant to discourage erratic shifts when assigning weights to component models as we move, week-by-week through an influenza season.

Our ensemble approach assumes a generative model for our data---ILI percentage at time $t$ ($i_{t}$)---by first sampling a component model $m$ with probability $\pi_m$, and second, sampling the ILI percentage from the component model $m$ with probability mass function $f_{m}(i)$.
(In practice, the submitted probabilistic forecasts $f_{m}(i)$ are probability mass functions defined over discrete bins of possible ILI incidence.
These are often discretized versions of probability densities.)
\begin{align}
  z(t) \sim \mathrm{Cat}\l(\pi_{1}, \pi_{2}, \cdots, \pi_{M}\r)\\
  i_{t} | z(t) \sim f_{z(t)}
\end{align}
where $z(t)$ is a vector of length $M$, each entry corresponding to one of $M$ possible component models, and $f_{z(t)}$ is the probability density the variable $z(t)$ selected to generate $i_{t}$, the ILI percentage at epidemic week $t$.
The variable $z(t)$ is an indicator variable and assigns the value $1$ when this model generated the ILI percentage $i_{t}$ and $0$ otherwise.
While a mixture model is likely not the true generating process for ILI data, this framework conveniently expresses the probability distribution over possible ILI values as a convex combination of component models by averaging over the random variable $z$.
This also makes clear the difference between combination forecasting and BMA.
Our combination forecast assumes there is a probabilistic distribution over which models produce ILI values at every time step~($z(t)$).
BMA would place a single probabilistic distribution over which model produced all ILI values.
We define the probability of observing an ILI percentage at time $t$,
\begin{equation}
  p(i_{t} | \pi) = \sum_{z(t)} p\l[z(t)\r] \times p\l[i_{t} | z(t)\r] = \sum_{m=1}^{M} \pi_m f_{m}(i_{t})
\end{equation}
and require the $\pi$s are non-negative and sum to one.
This mixture model framework for probabilistic model averaging is the standard for the field~\cite{reich2019collaborative,ray2018prediction,brooks2018nonmechanistic}.


\subsubsection{Expectation-Maximization method for weight estimation}
Though a convenient way to specify the model, directly optimizing the loglikehood is difficult.
The loglikelihood over all $T$ time points equals
\begin{equation}
    \log p(\mathcal{D} | \pi ) = \sum_{t=1}^{T} \log\l[\sum_{m=1}^{M} \pi_m f_{m}(i_{t})\r], \label{LL}
\end{equation}
and is a summation over log-sums, where $\mathcal{D} = [i_{1}, i_{2}, \cdots, i_{T}]$ is the observed $T$ ILI percentages.
An alternative method to optimize the loglikelihood~\cite{bishop2006pattern,murphy2012machine} considers the loglikelihood over both ILI percentages and the set of hidden indicator variables~$z(t)$---one for each week of data---that decide which component model (from $M$ possible models) was selected to generate data point $i_{t}$.
The hidden variables $Z$ simplify the likelihood to
\begin{equation}
  \begin{aligned}
    p(\mathcal{D},Z | \pi ) &= \prod_{t=1}^{T} \prod_{m=1}^{M} \l[\pi_m f_{m}(i_{t})\r]^{z(m,t)}\label{emLiklihood}
  \end{aligned}
\end{equation}
where $z(m,t)$ equals $1$ when the $m^{\text{th}}$ model generated the $t^{\text{th}}$ ILI observation and $0$ otherwise, and $Z$ is a $M \times T$ matrix of all hidden variables.
The Expectation-Maximization (EM) algorithm~\cite{dempster1977maximum,reich2019collaborativepnas} iteratively optimizes a lowerbound on the logliklihood, the expectation over $p(\mathcal{D},Z | \pi )$ with respect to $p(Z | \mathcal{D}, \pi^{t-1})$ or $\mathbbm{E}_{p(Z | \mathcal{D}, \pi^{t-1})} p(\mathcal{D},Z | \pi )$ to find the component model weights that maximize the loglikelihood~(\textbf{deEM-MM} Fig.~\ref{algorithms} left).
The prefix \textbf{de} (degeneracy) refers to this algorithm's inability to change parameters inside component models and the suffix \textbf{MM} (mixture model) references the assumed data generating process.

\subsubsection{Variational Inference for weight estimation}

While the deEM-MM algorithm find an optimal set of weights, it considers $\pi$ a fixed vector of parameters.
Assuming $\pi$ is a random variable, we can restructure our problem as Bayesian and infer a posterior probability over weights.
Our Bayesian framework extends~\eqref{emLiklihood} by modeling the posterior probability of $\pi$
\begin{equation}
  p(\pi | Z, \mathcal{D}) \propto p(\pi) \times p(\mathcal{D},Z | \pi).
\end{equation}
While many different choices are available for a prior over ensemble weights, $p(\pi)$, we chose a Dirichlet distribution because it is conjugate to the likelihood.
We further decided on a Dirichlet prior that has a single parameter~($\alpha$) shared across all $M$ models. 
The scale of the shared parameter~($\alpha$) governs the influence of the prior distribution on the posterior.
A larger value of $\alpha$ shrinks ensemble weights closer towards an equally weighted ensemble.
Equal weights are a reasonable first choice for many ensemble settings.
Smaller values allow the model to rely more heavily on the observed data. 

Typically the prior is fixed during training. 
However, in our setting, where the amount of data is growing and recent past ILI percentages will be revised as we move through a season, we allow the prior to be time-dependent.
A time-dependent prior can act to continually weaken the influence of the (likely to be revised) data on the ensemble weights and consistently pull component model weights towards equal over the course of the season.

We specify the prior as 
\begin{align*}
    \pi_{t} &\sim \text{Dir} \l[\alpha(t)\r]\\
    p[\pi_{t} | \alpha(t)] &= \f{\Gamma \l[\sum \alpha(t) \r]}{\prod \Gamma\l[\alpha(t)\r]} \prod_{m=1}^{M}\pi_m^{\alpha(t)-1}
\end{align*}
where $\alpha(t)$ is the parameter (the same across all $M$ component models) that defines the Dirichlet distribution at time $t$.
We chose $\alpha(t)$ to be a constant fraction~($\rho$) of the number of training data points for fitting at time $t$ divided equally amongst all $M$ component models
\begin{equation}
  \alpha(t) = \rho \f{N(t)}{M},
\end{equation}
where $N(t)$ is the number of training data points, $M$ is the number of component models, and $\rho$ is a value between $0$ and $1$.
Throughout the season the prior parameter will grow linearly as the ensemble trains on an increasing number of observations and regularize component model weights at a constant rate.

We can plug our specific prior in for time $t$
\begin{equation}
  \begin{aligned}
    p(\pi_{t} | Z, \mathcal{D}) &\propto p\l(\pi_{t}\r) \times p\l(\mathcal{D},Z | \pi_{t}\r)\\
    &= \text{Dir}\l[\pi_{t} | \alpha(t)\r] \times p\l(\mathcal{D},Z | \pi_{t}\r) \label{bayesModel}.
    %
  \end{aligned}
\end{equation}
and find an optimal set of mixture weights by optimizing the following function of $\pi$ and $Z$
\begin{align}
    \log p\l(\pi_{t}|Z,\mathcal{D}\r) \propto \sum_{t=1}^{T} \sum_{m=1}^{M} \mathbbm{1}\l[z(m,t)\r] \log\l[\pi_{mt}f(m,i_{t})\r] + \sum_{m=1}^{M} \l[\alpha(t)-1\r] \log \pi_{mt}. \label{eq:logposterior}
\end{align}
We note that the first term on the right-hand side of equation (\ref{eq:logposterior}) is the same as taking the log of the frequentist expression for the likelihood shown in equation (\ref{emLiklihood}).
Intuitively, we expect a stronger (increasing $\rho$) even prior to encourage weights across components to move towards $1/M$, uniformly distributing probability mass to every component model.

To estimate the posterior probability over ensemble weights~\eqref{eq:logposterior},  We will use a variational inference algorithm~(\textbf{deVI-MM}).
Variational inference is similar to the EM algorithm and both can be motivated by introducing a distribution over hidden variables and decomposing the loglikehood into two parts: a lowerbound $\mathcal{L}$ on the log likelihood $\log p(\mathcal{D}|\pi)$ and an error term that is the difference between the lowerbound and loglikehood~$\l( \mathcal{L} - \log p\l(\mathcal{D}|\pi \r) \r)$.

Given weights $\pi$, hidden variables $Z$, and fixed data $\mathcal{D}$, we first rewrite the posterior over $Z$ and $\pi$
\begin{equation*}
  p(Z,\pi | \mathcal{D}) = \f{ p(Z, \pi, \mathcal{D}) }{p(\mathcal{D})}
\end{equation*}
Reordering terms and taking the log,
\begin{equation*}
  \log \l[ p(\mathcal{D}) \r] = \log \l[  \f{p(\mathcal{D}, Z,  \pi)}{p(Z, \pi | \mathcal{D})}\r] = \log \l[ \f{p(\mathcal{D} | Z,\pi) p(Z,\pi) }{ p(Z,\pi | \mathcal{D})} \r] \nonumber.
\end{equation*}
We can express the marginal loglikelihood in terms of the complete loglikelihood (numerator) and log of the posterior distribution. 
The last step introduces a distribution over hidden variables $Z$ and $\pi$
\begin{equation}
  \log \l[ p(\mathcal{D} ) \r] = \log \l[ \f{p(\mathcal{D} | Z,\pi) p(Z,\pi) }{q(Z,\pi)} \times \f{q(Z,\pi)}{ p(Z,\pi | \mathcal{D})} \r] \nonumber
\end{equation}
and integrates over $q(z)$
\begin{align}
  \log \l[ p(\mathcal{D}) \r] &= \mathbbm{E}_{q} \l\{ \log \l[ \f{p(\mathcal{D} | Z,\pi) p(Z,\pi)}{q(Z,\pi)}  \r]  \r\} + \mathrm{KL} \l[q(Z,\pi) \| p(Z,\pi| \mathcal{D}) \r] \label{twoTerms}
\end{align}
where $\mathrm{KL}$ is the Kullback-Leibler divergence~\cite{cover2012elements}, a non-negative function of $q$.
Since $\mathrm{KL}$ is a non-negative function, the first term on the right hand side is a lower bound to the marginal loglikeihood and the second term, the $\mathrm{KL}$ term, is the difference between the marginal loglikelihood and the lower bound.
Iteratively optimizing this lower bound can be shown to monotonically improve the loglikelihood and converge on the best possible representation ($q$) of the complete loglikelihood. 

The EM and VI algorithms diverge on how to choose $q$: the EM algorithm considers $\pi$ a fixed set of parameters and chooses $q = p(Z|\pi,\mathcal{D})$, zeroing out the above Kullback-Leibler divergence but also assuming $p(Z | \pi, \mathcal{D})$ can be computed~(see~\cite{bishop2006pattern} for theoretical development and \cite{reich2019collaborativepnas} for an application to infectious disease forecasting).
The VI algorithm allows any choice for $q$.
The mean-field approximation, the most common choice for $q$, factors $q$ into discrete pieces $q = \prod_{j}q_{j}(z)$~\cite{blei2017variational,rustagi1976variational} and is the approach we take.
We chose to factor our distribution over hidden variables into a distribution over $Z$ and separate distribution over $\pi$, $q(\pi,Z) = q(Z) \cdot q(\pi)$, dividing indicator variables from mixture weights.

This choice yields (see details in Suppl. \ref{sec:appendixA}) a $q(\pi)$ that follows a Dirichlet distribution and indicator variable $z(m,t)$ that follows a Bernoulli distribution:
\begin{align}
    q(\pi)  &\sim \mathrm{Dir}\l( \alpha(t) + \sum_{t=1}^{T}  r(m,t)  \r) \label{dirSol}\\
   q\l[z(m,t)\r] &\sim \mathrm{Bern}\l[ r(m,t) \r],
\end{align}
where the responsibility $r(m,t)$, or the probability that model $m$ generated datapoint $i_{t}$, equals
\begin{equation}
  r(m,t) = \dfrac{ \exp \l \{ \mathbbm{E}_{\pi} \log \l[ \pi_m \r] + \log \l[ f_{m}(i_{t}) \r] \r \} } {\sum_{m=1}^{M} \exp \l \{ \mathbbm{E}_{\pi} \log \l[ \pi_m \r] + \log \l[ f_{m}(i_{t}) \r] \r \}}
\end{equation}
and the responsibilities summed over component models $(m)$ equals one.
The distribution $q\l[z(m,t)\r]$ computes the probability ILI value $i_{t}$ was generated by model $m$.

The Variational approach here can also be interpreted as a generalization of the EM solution.
If we fix $\pi$ to be a non-random parameter, the responsibilities reduce to
\begin{equation}
  r(m,t) = \dfrac{\pi_mf_{m}(i_{t})}{\sum_{m=1}^{M} \pi_mf_{m}(i_{t}) },
\end{equation}
the same responsibility function for the EM algorithm~\cite{reich2019collaborativepnas}.
Both VI and EM algorithms will find a unique global maximum of the log-likelihood~\eqref{LL} due to the convexity of the log-likelihood function~(see analysis of convexity in Suppl.~\ref{convexity}).

\subsubsection{How prior impacts ensemble weights}
The prior can be recognized as a regularizer.
The maximum aposteriori estimate (MAP) for the $m^{\text{th}}$ component model can be computed  computed directly by dividing the weight given to model $m$ by the sum over all weights (sum over $m$)
\begin{equation*}
  \text{MAP}\l[\pi_m\r] = \f{ \alpha(t) + \sum_{t=1}^{T}r(m,t)}{ \sum_{m}\alpha(t) + N(t) }
\end{equation*}
where $\sum_{m,t}r(m,t)=N(t)$ is the total number of ILI values used for training at time $t$.
Though useful for computation, an alternative characterization expresses the MAP as a convex combination of the prior and the percent responsability assigned to component model $m$.
Define the prior percent weight assigned to component model $m$ as $\alpha_{m}(t)/\sum_{m}\alpha_{m}(t)$ and the percent responsibility assigned to component model $m$ due to the data as $\sum_{t}r(m,t)/N(t)$.
The MAP of $\pi$ can be reexpressed as a convex combination of the prior responsibility plus responsibility learned form the data
\begin{equation}
  \text{MAP}\l[\pi_m\r] = \l[ \f{\alpha(t)}{\sum_{m} \alpha(t)}\r]\l(\f{\rho}{1+\rho}\r) + \l(\f{1}{1+\rho}\r) \l[\f{\sum_{t=1}^{T}r(m,t)}{N(t)}\r].
\end{equation}
Strengthening the prior, in our case increasing $\rho$, shifts the MAP estimate away from the data-estimated responsibility and towards the prior.
The constant shift, despite increased training data, is a result of our time-dependent prior.
By setting our prior percentage to equal across all models in the ensemble $(1/M)$, the prior weight $\rho$ interpolates between an equally-weighted ensemble (ignoring past model performance) and a completely data-driven weighting.



\subsubsection{deEM-MM and deVI-MM Algorithms}
A comparison of the EM and VI algorithms is shown below in Fig.\ref{algorithms}.
The EM and VI algorithms are similar to one another: both rely on adding hidden variables to the loglikelihood and iteratively maximizing a lower bound.
When computing $Z$, both algorithms need the probability that component model $m$ generated data point $i$ for models $1$ to $M$ and data points $1$ to $T$. 
A key difference is how the EM and VI algorithms approximate the distribution over hidden variables $Z$.
The EM algorithm requires the previous estimate of ensemble weights, but the VI algorithm considers the weights a random variable and needs to compute the ensemble weight's expectation~(see Suppl.~\ref{app.expec} for details on calculating $\mathbbm{E}[\pi]$).
These differences are present in step $7$, updating $Z$, and step $10$ updating $\pi$.
Another difference is in evaluating model fit.
The EM algorithm computes the loglikelihood, log of \eqref{emLiklihood}, while the VI algorithm computes a related quantity called the Evidence Lower Bound (ELBO).
The ELBO is defined as 
\begin{equation}
    \text{ELBO} = \text{loglikelihood} - \l\{\log[q(\pi)] + \log[q(Z)]\r\},
\end{equation}
or the difference between the loglikelihood and our approximation over hidden variables.
When updating the ensemble weights, both algorithms sum over the probability $i_{t}$ belongs to model $m$, but the VI algorithm adds an additional term to the ensemble weight, the prior.

\centerline{
\begin{minipage}[th]{8cm}
  \begin{algorithm}[H]
    \label{deEM}
    \caption{deEM-MM Algorithm}
    \begin{algorithmic}[1]
      \State input: $i_{1 \times T}$, $\pi_{0}$, $\tau$
      \State output: $\pi$
      \State
      \State $\ell\ell \gets []$
      \State$\pi_{M \times 1}$ = $\pi_{0}$
      \For{j=1:maxIters}
      \State $Z_{M \times T} \gets \pi \times f(i)$
      \State $Z \gets Z/\text{colSum}(Z)$
      \State $\pi \gets$ rowSum($Z$) 
      \State $\pi \gets \pi/$ sum($\pi$)
      \State $\ell\ell[j] \gets$ computeLL($i$,$\pi$)
      \If{ LL[j] - LL[j-1] $<$ $\tau$}
      \State break
      \EndIf
      \State return $\pi$
      \EndFor
  \end{algorithmic}
\end{algorithm}
\end{minipage}
\begin{minipage}[th]{8cm}
  \begin{algorithm}[H]
    \label{deviMM}
    \caption{deVI-MM Algorithm}
    \begin{algorithmic}[1]
      \State input: $i_{1 \times T}$, $\pi_{0},\alpha_{M \times 1}$, $\tau$
      \State output: $\pi$
      \State
      \State $\text{ELBO} \gets []$
      \State$\pi_{M \times 1}$ = $\pi_{0}$
      \For{j=1:maxIters}
      \State $Z_{M \times T} \gets \exp\l( \mathbbm{E}\l( \log \pi\r) + \log f(i)\r)$
      \State $Z \gets Z/ \text{colSum}(Z)$
      \State $\pi \gets$ rowSum($Z$) 
      \State $\pi \gets \pi/$ sum($\pi$) + $\alpha$
      \State $\text{ELBO}[j] \gets$ computeELBO($i$,$\pi$)
      \If{ ELBO[j] - ELBO[j-1] $<$ $\tau$}
      \State break
      \EndIf
      \State return $\pi$
      \EndFor
    \end{algorithmic}
  \end{algorithm}
\end{minipage}
}
\captionof{figure}{
  An Expectation maximization~(left) and Variational Inference~(right) algorithm estimating ensemble weights for a Frequentist and Bayesian adaptive algorithm.
  Although both algorithms have different underlying probability models, they follow similar steps.
  The major differences between EM and VI algorithms are: how $Z$ is computed (step $7$), the prior over $\pi$ (step $10$), and how weights are evaluated (steps $11$ and $12$).
  ~\label{algorithms}}

\subsection{Experimental design}

Five ensemble models will be analyzed in detail~(Table.~\ref{tb.models}).
The equally-weighted ({\bf EW}) ensemble assigns the same weight $(1/M)$ to all component models.
The {\bf Static} ensemble trains on all past component model forecasts of ILI and assigns weights to component models before the start of the next season.
Weights are kept fixed throughout the next season.
The adaptive model with three types of regularization, that correspond to three values of $\rho$, will be studied: close to 0\% regularization ({\bf Adaptive$_{\text{non}}$}), `optimal' regularization ({\bf Adaptive$_{\text{opt}}$}), and `over' regularization ({\bf Adaptive$_{\text{over}}$}).
We will consider our adaptive model `optimally' regularized by computing logscore statistics for the held-out $2010/2011$ season and choosing the prior that has the highest average logscore.

\begin{table}[ht!]
    \footnotesize
    \begin{tabular}{lL}
        \hline
        Model & Description\\
        \hline
        Equally-weighted (EW)  & Assign a weight of $1/21$ to every component model               \\
        Static with no regularization (Static) & Weights are trained on all previous seasons and kept fixed throughout the target season \\
        Adaptive with no regularization (Adaptive$_{\text{non}}$) & Weights are trained within season, change week to week, and have a 0\% prior regularization. \\
        Adaptive with optimal regularization (Adaptive$_{\text{opt}}$) & Weights are trained within season, change week to week, and have a optimal prior calculated from the 2010/2011 season.\\
        Adaptive with over regularization (Adaptive$_{\text{over}}$) & Weights are trained within season, change week to week, and have a prior greater than the optimal\\
        \hline
    \end{tabular}
    \caption{Description of the $5$ ensemble models analyzed.\label{tb.models}}
\end{table}

\subsubsection{Training and scoring component models for ensembles}

Component models were scored on unrevised ILI data observed throughout each season.
We created a record of the component model scores, as if they had been scored in real time, every week throughout each season.
Adaptive models were trained on component model logscores throughout a season and this impacted how the adaptive model was trained in two ways: (i) every epidemic week new ILI data was observed and generated new component model logscores, and (ii) past ILI data was revised, which in turn changes past component model logscores.
Our static ensemble was trained on component model log scores from past seasons, not the current season.
At the end of a season ILI data is finalized and component model log scores are final, and so training data was constant for the static ensemble.
Ensemble log scores were calculated on ILI percentages reported on EW28, the ``final'' week used by the CDC FluSight challenges.
ILI values at or after EW28 for each season are considered definitive.
Based on previous work~\cite{reich2019collaborativepnas}, and due to the relatively smaller amount of data available to the adaptive ensemble, we chose to fit a  `constant weight' ensemble model for both the static and adaptive approaches.
The constant weight model assigns one weight to each component model, the weight does not vary by region or target.

\subsubsection{Fitted ensemble models}

We computed adaptive ensembles for prior values from 1\% to 100\% by 1\% increments for the $2010/2011$ season.
To determine a prespecified `optimal prior', we chose the prior corresponding to the highest average logscore.
The $2010/2011$ season was removed from all formal comparisons between ensembles.
The adaptive ensemble corresponding to the optimal prior was used for formal comparisons.

For each season in $2011/2012$ through 2017/2018, we computed \adaptNon, \adaptOpt, \adaptOver,  ensembles.
Static ensemble and equally-weighted ensembles were also fit at the beginning of each season, using only data for prior seasons.

\subsubsection{Formal Comparisons}

To compare ensemble models, we computed the difference between the logscore assigned to one ensemble versus another.
A random intercept model will describe the difference between logscores, averaged over epidemic weeks and paired by season, region, and target.
\begin{align}
    D_{w,s,r,t} &\sim \mathcal{N}\l( \beta_{0} + a_{s} + b_{r} + c_{t}, \sigma^{2} \r)\\
    a_{s} &\sim \mathcal{N}\l(0,\sigma^{2}_{a}\r) \nonumber \\
    b_{r} &\sim \mathcal{N}\l(0,\sigma^{2}_{b}\r) \nonumber \\
    c_{t} &\sim \mathcal{N}\l(0,\sigma^{2}_{c}\r) \nonumber
\end{align}
where $D$ is the difference between logscores assigned to two ensembles: for epidemic week $w$, in season $s$, region $r$, and for target $t$.
The fixed intercept (average difference in logscore between ensemble models) is denoted $\beta_{0}$, and season ($a_{s}$), region ($b_{r}$), and target ($c_{t}$) effects are Normally distributed ($\mathcal{N}$) with corresponding variances.

We will fit two random effects models: one comparing the adaptive vs. equally-weighted ensemble, and the second comparing the adaptive vs. static ensemble.
The conditional mean, $95$\% confidence interval, and corresponding p-value will be reported.
An additional bootstrapped p-value is also reported.
Random samples (with replacement) are selected from the set of all season-region-target-epidemic week tuples.
For every random sample, a random effects models is fit and conditional means collected for: seasons, regions, and targets. 
The set of random samples is centered to create a null distribution and compared to our original dataset's conditional mean.
We report the empirical probability a random sample from the centered null distribution exceeds our original dataset's conditional mean.

\section{Results}

\subsection{Choosing prior to maximize performance}

\begin{figure}[ht!]
  \centering
  \includegraphics[scale=0.60]{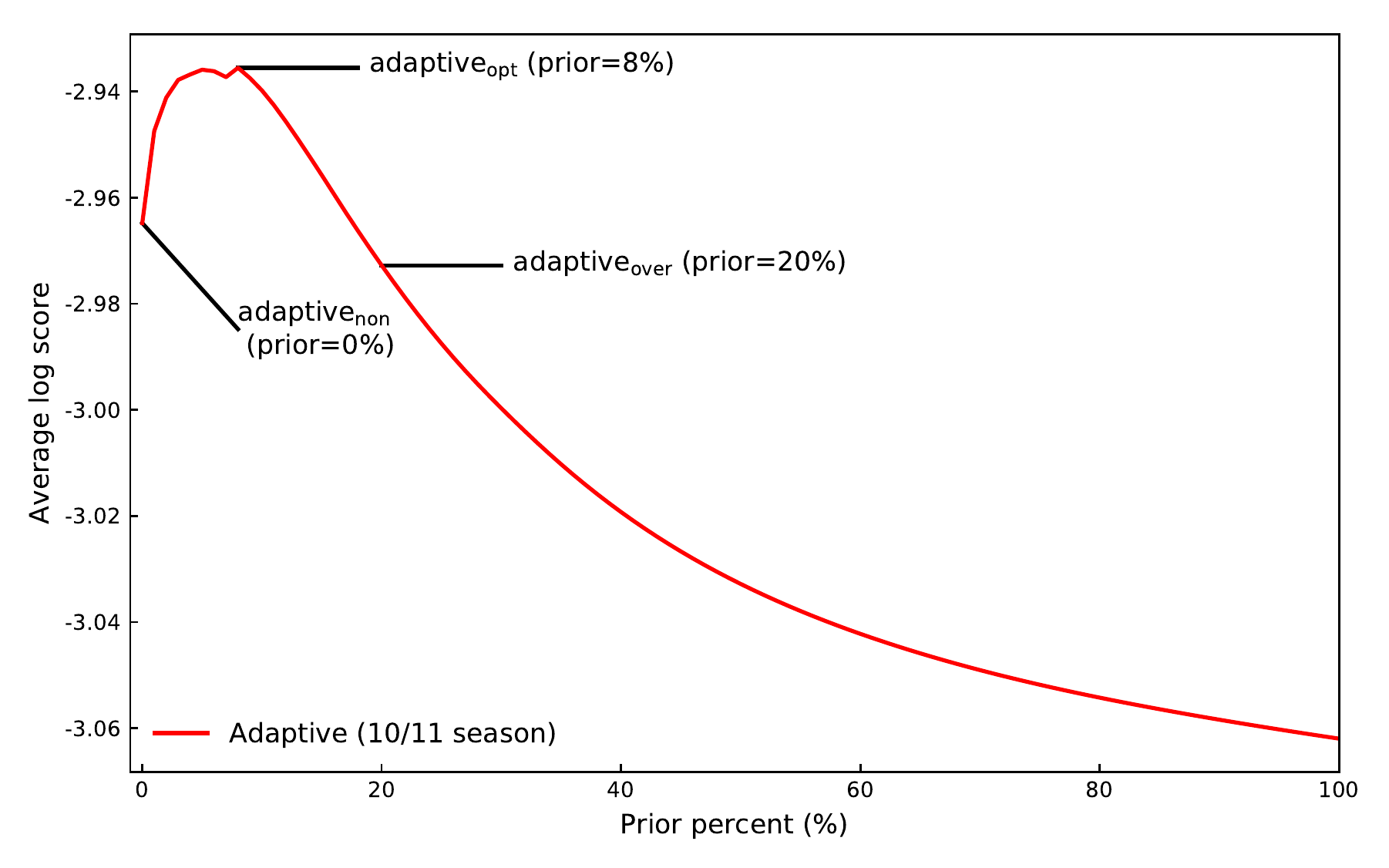}
  \caption{ The logscore for the 2010/2011 season, averaged over region, target, and epidemic week for priors from 0\% to 100\% by 1\% increments.
   The prior corresponding to the maximum logscore~(prior=8\%) was chosen as our \adaptOpt ensemble for formal comparisons to static and equally weighted ensembles. \label{fig2.prespecifiedLogScorePlot}}
\end{figure}

We chose an optimal prior for the adaptive model by fitting our adaptive model to the $2010/2011$ season for priors from 0\% to 100\% by 1\% and selecting the prior with the highest average logscore~(Suppl.~\ref{suppl1.logScoresPerSeason} includes adaptive fits and average logscores for all seasons from $2010/2011$ to $2017/2018$).
The average logscore for the $2010/2011$ season peaks at a prior of $8$\%~(Fig.~\ref{fig2.prespecifiedLogScorePlot}), and we will consider an adaptive model with $8$\% prior~($\rho=0.08$) our \adaptOpt ensemble.
After a prior of $8$\%, the logscore sharply decreases.
This decrease in performance suggest ensemble weights are over regularized, and we chose a $20$\% prior as our \adaptOver ensemble.
Finally, a $0$\% prior was chosen as our \adaptNon ensemble.

\subsection{Prior successfully regularizes ensemble weights} 

\begin{figure}[ht!]
    \centering
    \includegraphics[scale=0.70]{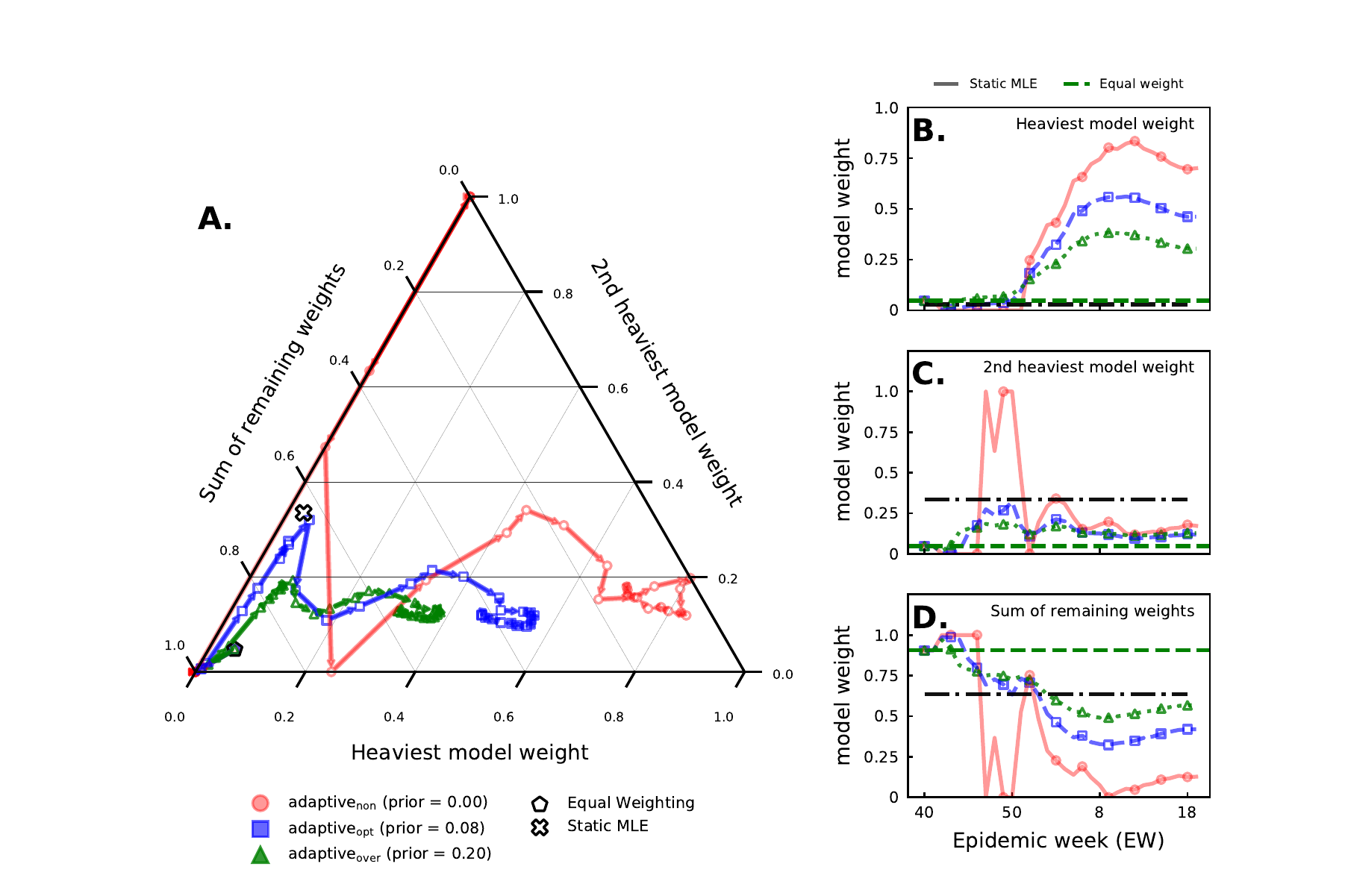}
    \caption{
    Component model weights plotted over epidemic weeks all ensembles during the 2017/2018 season. 
    We highlight the weights for the two component models the \adaptNon ensemble assigned the most weight to at the end of the season and compare this to the sum of all other model weights. 
    (A) A simplex plot shows how, for each
    ensemble, the weights assigned to the top two models and the rest of the models move together across the season. 
    The lines can be seen as a trajectory, beginning at the black pentagon (the equal-weight ensemble) and, following the arrows, migrate through the simplex with each point representing the triple $(\hat\pi_t^{(1)}, \hat\pi_t^{(2)}, \sum_{j=3}^M\hat\pi_t^{(j)})$ of weights estimated in week $t$. The estimated weights from the static ensemble using all data prior to 2017/2018 (but, unlike the adaptive ensemble, using no data from this season) are represented by the `x'.
    Plots of $\hat\pi_t^{(1)}$ (panel B) and $\hat\pi_t^{(2)}$ (panel C) across all weeks $t$ in the 2017/2018 season. The estimates of this model's weight from the static ensemble and the equal weight (1/21) are shown in horizontal dashed lines.
    (D) A plot of $\sum_{j=3}^M\hat\pi_t^{(j)}$ across weeks $t$. The estimates of the sum of these models' weights from the static ensemble and the equal weight (19/21) are shown in horizontal dashed lines.
     \label{fig3.dorito}}
\end{figure}

We compared the \adaptNon, \adaptOpt, and \adaptOver ensembles ($\rho$ = 0.01, 0.08, and 0.20 respectively) to investigate how the ensemble weights change with the prior.
Adding a prior regularizes component model weights~(Fig.~\ref{fig3.dorito}).
Smaller priors yield higher variability in ensemble weights throughout any given season. 
For example, in 2017/2018~(Fig.~\ref{fig3.dorito}), this is especially evident for the \adaptNon ensemble, when ensemble weights vacillate between assigning almost all weight to one component model or another.
The \adaptOpt and \adaptOver weights, in comparison, do not show as high variability over time.
Component model weights for all three ensembles do track one another, with specific model weights moving up and down in unison, albeit in a more muted fashion for the stronger 8\% prior (\adaptOpt) and 20\% prior (\adaptOver).
The patterns shown in~Fig.~\ref{fig3.dorito} persist in other seasons as well (see Suppl.~\ref{simplexplot}).

\subsection{Comparing adaptive vs.~equally-weighted and static ensembles}

\begin{figure}[ht!]
  \centering
  \includegraphics[scale=0.75]{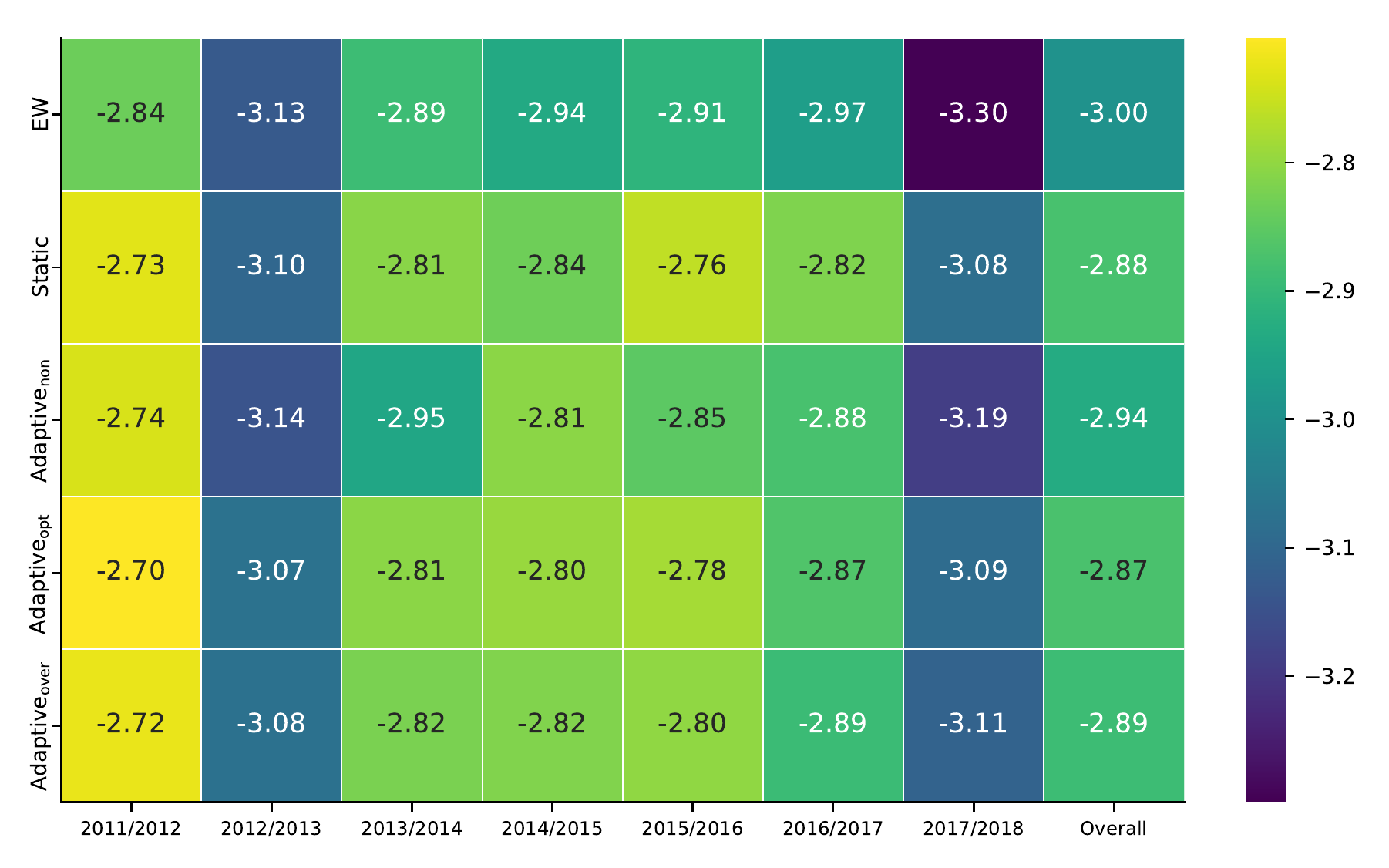}
  \caption{
    Logscore averaged over region and target for ensemble algorithms stratified by season.
    The 8\% prior was pre-specified as the optimal prior based on analysis of the excluded 2010/2011 season.
    All ensembles show variability in performance across seasons. 
    Adaptive and static ensembles outperform the equal weighted ensemble, the adaptive and static ensemble showing similar performance.
    Adaptive ensembles perform similar to static ensembles despite having less training data in later seasons.
   \label{fig4.logScoresGrid}}
\end{figure}

Adaptive ensembles consistently outperform equally-weighted ensembles, and show comparable performance to static ensembles~(Fig.~\ref{fig4.logScoresGrid}).
The \adaptOpt model has higher logscores than either \adaptNon and \adaptOver models, and the EW model.
The \adaptOver, unlike the \adaptNon model, always outperforms the EW model, indicating it is better to over- than under-regularize, at least to some degree.
The static model---trained on multiple seasons of data---performs the best. 
Adaptivity improves over assigning equal weights to component models, and performs similar to, but cannot outperform, the data-rich static model. 

\begin{figure}[ht!]
  \centering
  \includegraphics[scale=0.75]{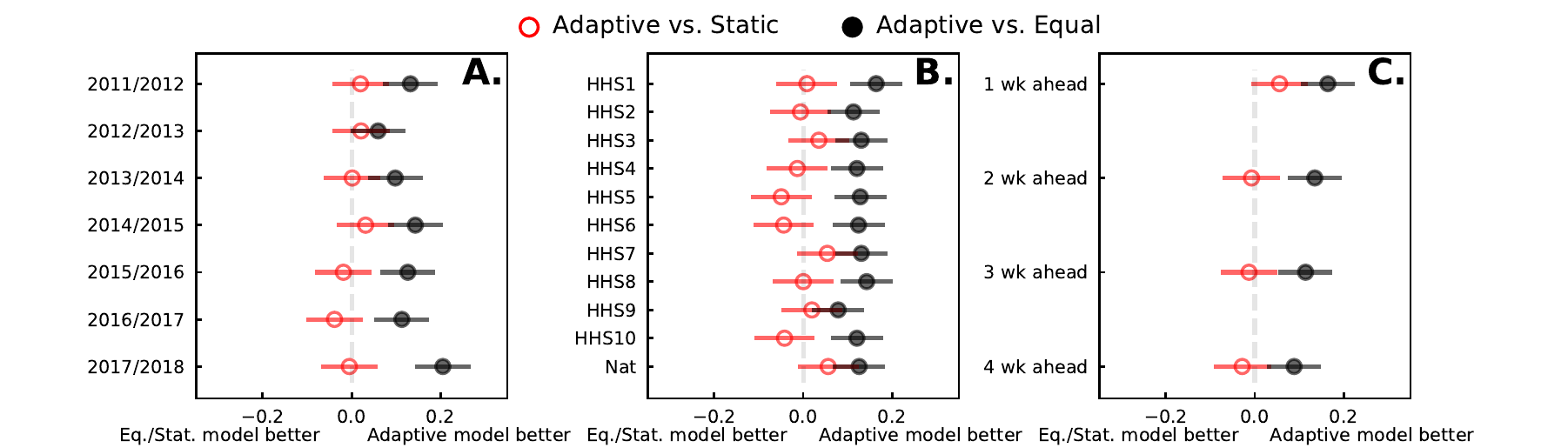}
  \caption{
    Logscores from equally-weighted, static, and adaptive ensembles are within season, region, target, and epidemic week.
    Estimates and 95\% confidence intervals are plotted from two random intercepts models: modeling the difference in logscore between adaptive and equally-weighted ensembles and the second the difference between adaptive and static ensembles.
    Positive differences suggest the adaptive ensemble performs better than the reference ensemble.
    Adaptive ensembles outperform the equally-weighted ensembles.
    Equally-weighted and adaptive ensembles do perform similarly in the 2012/2013 season.
    Adaptive ensembles perform similar to static ensembles.
    Differences in logscores favoring static or adaptive ensembles cannot obtain statistical significance for both parametric and bootstrapped p-values.
    Adaptive ensembles consistently outperform equal-weighted ensembles, and show similar or slightly improved (but not significant) performance against static ensembles.
    \label{fig5.pairedLogScoreDifferences}}
\end{figure}

Formal comparisons~(see Fig.~\ref{fig5.pairedLogScoreDifferences} and regression table in Suppl.~\ref{tbRegress}) demonstrate the \adaptOpt ensemble has statistically higher logscores compared to the EW ensembles, and perform similarly to static ensembles.
The \adaptOpt ensemble does perform worse against the static models for any particular choice of season, region, and target.
EW and \adaptOpt ensembles perform similar for the 2012/2013 season.
The 2015/2016, 2016/2017, and 2017/2018 season does show better (but not significant) performance for the static ensemble.
This suggests that the static model's performance comes from training on multiple seasons of data.
Differences in performance are less variable when stratified by region~(see Fig.~\ref{fig5.pairedLogScoreDifferences}B).
Difference in logscores between all ensembles decrease as forecasting horizons increase.
This reflects the difficulty in forecasting further into the future rather than ensemble performance.

In select strata the adaptive ensemble shows a statistically significant higher logscore compared to the EW and close to signficiant difference compared to the static ensemble. 
This significance is not enough to conclude the adaptive ensemble should outperform the static ensemble in most cases, however the data suggest, unsurprisingly, that the static ensemble performs relatively better the more data is has to train on.

\subsection{The adaptive ensemble's ability to learn seasonal variation}

\begin{figure}[ht!]
    \centering
    \includegraphics[scale=0.70]{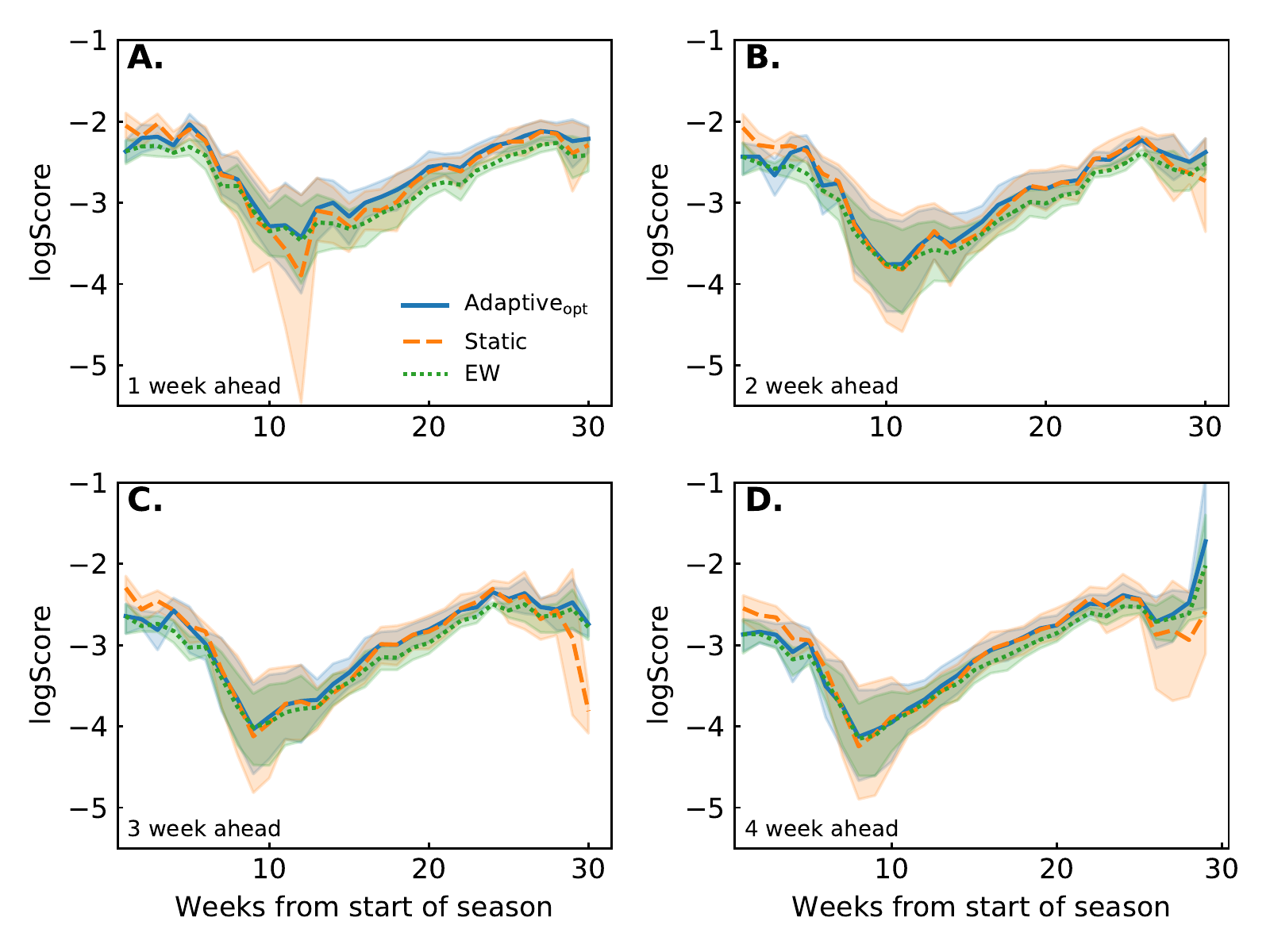}
    \caption{
    Average logscore over epidemic weeks stratified by target for adaptive$_{\mathrm{opt}}$ (blue solid line), static (orange dashed line), and EW (green dotted line) ensembles.
    Across targets, all models perform better at the beginning and end of the season.
    Performance decreases at approximately 10 weeks into the season, corresponding with peak ILI values.
    Comparing ensembles, the adaptive$_{\mathrm{opt}}$ model performs better than the EW model at all time points, the difference between performance increasing throughout the season.
    At the beginning of the season the adaptive$_{\mathrm{opt}}$ model performs worse than the static model, performing similar or better within 10 weeks. 
    The $1$ week ahead target shows the adaptive$_{\mathrm{opt}}$ model outperforming the static model by week $10$.
    The remaining targets show the adaptive$_{\mathrm{opt}}$ model performing similar to the static.
    \label{fig6.logScoresOverEpidemicWeeks}}
\end{figure}

The adaptive model has the opportunity to outperform the static model by tailoring component model weights within season.
Training weights based on component model performance data throughout the season is useful, as is shown by the \adaptOpt model outperforming the EW model.
But the adaptive ensemble must accrue several weeks of training data before it can perform similar, and in some cases better, than the static ensemble~(Fig.~\ref{fig6.logScoresOverEpidemicWeeks}).
When this increase in adaptive performance occurs highlights two key points: (i) the adaptive ensemble is learning seasonal variation and (ii) the static model's better performance early in the season suggests some continuity between component model weights at the end of one season and beginning of the next.

However, the adaptive models also are able to, without any prior data, learn how to create an optimal weighting within a season. 
While the adaptive models may be penalized early on in a season by not possessing the historical knowledge of which models have performed well, they adjust for this by learning which models are performing well in a particular season. 
This is illustrated clearly in Fig.~\ref{fig3.dorito} where the heaviest model at the end of the season (with 46.2\% of the weight for the \adaptOpt ensemble) was assigned only 4.82\% of weight at the beginning of the season. 
This ability to adapt to within-season dynamics appears to provide the \adaptOpt model with a slight advantage over the static model in the middle and end of the season.

\section{Discussion}

We developed a new algorithm for assigning weights to component models.
Our model extends a previous ensemble algorithm~\cite{reich2019collaborative} by assuming component model weights are random variables, rather than fixed quantities.
This model reassigns component model weights every week, but influenza data is prone to change within a season.
For settings with sparse and possibly revisable data, we introduced a time-dependent uniform Dirichlet distribution that regularizes ensemble weights. 
This adaptive ensemble outperforms an equally-weighted ensemble, and performs similarly to a static ensemble that requires multiple years of data to perform well.

It is expected that learning from training data, and the amount available, would contribute to better predictive performance.
The equally-weighted model ignores any training data, and similar to an adaptive ensemble with no training data, assigns equal weights.
The equally-weighted ensemble performs worst.
The next best model, the adaptive ensemble, trains on component model performance within the same season.
Even though the static model has multiple seasons of training data available, and does have the best performance, the adaptive ensemble is not far behind.
It is difficult to tell whether the similar performance is a shortcoming of the static or adaptive ensemble.
Similar performance between ensembles could be because component model performance from past seasons (available to the static ensemble) does not generalize to future seasons.
Alternatively, the adaptive ensemble may not be using the within-season training data efficiently and could be leaving behind hidden patterns in the component model forecasting data. 
But we did find static weights outperforming the adaptive ensemble early on in the season.
These results suggest a model could perhaps use the static ensemble weights, when available, as a prior for the adaptive ensemble.

Regularizing ensemble weights increased adaptive ensemble performance.
Not only can regularization improve adaptive model performance, but could also improve the performance of the static model~(Suppl.~\ref{fig.staticHelp}).
Improving performance by regularizing ensemble weights with a time-dependent uniform Dirichlet prior may be applicable to any ensemble weighting scheme. 

Though our empirical work suggests an optimal prior near 8\%, we expect the optimal prior to vary by season, region, and target.
Different priors will likely be optimal for different biological phenomena too.
A more flexible prior could improve adaptive ensemble performance:
changing the prior percent throughout the season, allowing the prior to depend on performance data from past seasons, or modeling the prior on regional and target information.
Instead of limiting the prior as a regularizer, we can include outside information from past seasons, or patterns in influenza data the ensemble cannot learn from training data.
Future research will explore different methods of modeling prior ensemble weights.

Our work here connects to Gneiting and other's work on linear pooling~\cite{ranjan2010combining,gneiting2013combining,geweke2011optimal,wallis2011combining}.
Linear pools assume, unlike Bayesian model averaging, that the data is generated by a combination of component models.
Similar to Gneiting's work, our weights are optimized with respect to the loglikelihood of a generative model, but the loglikelihood is really a different representation of a logscore---the log of the probability corresponding to weights $\pi$.
Our model could be fit by minimizing a generic loss function based on logscore, but we found the EM (and VI) algorithms fit our model fast and consistently found a global optimum. 

Unlike Gneiting, we did not recalibrate the combined predictive distributions, and methods like the beta transformed linear pool~\cite{ranjan2010combining} or shifted linear pool~\cite{kleiber2011locally} could improve predictive performance even further.

This paper has many limitations future work can address: (i) better choice of prior, (ii) accounting for correlated component models, (iii) post-processing or `recalibrating' our combined forecast.
Our adaptive ensemble examined how a prior can impact ensemble performance, but we only explored a uniform prior.
Future work could study informative priors and how to manipulate priors during the course of a season to improve ensemble performance.
Our model also assumed component models made independent forecasts.
A more advanced ensemble would examine the correlation structure of the data~(region, target), and the component model forecasts.
Finally, our ensemble model focused on an optimal set of weights for forecasting, and made no efforts to recalibrate the combined forecast.\cite{gneiting2013combining}

From a public health perspective, officials have been moving away from stand-alone subjective decision making in favor of incorporating data-driven analyses into decisions.
This trend has become particularly apparent in infectious disease outbreak management.\cite{Rivers2019}
Adaptive ensembles, providing real-time forecasts of infectious disease, supports this trend towards evidence based decision making for public health.
Combining their expertise with statistical models like our adaptive ensemble, public health officials can work towards impeding outbreaks and changing public health policy.
Public health officials often find themselves making decisions in the middle of crises, times when fast effective forecasting is necessary. 
The adaptive ensemble's flexibility, reliance only on near-term data, and capacity to track unusual disease trends can support accelerated public health decision making.

Adaptive ensembles---reweighting component models, week by week, throughout a season---can handle sparse data scenarios, admit new component models season by season, and shows similar prescience compared to the more data-heavy static ensemble.
This added flexibility makes them an important tool for real-time, accurate forecasts.

\section{Acknowledgements}
This work has been supported by the National Institutes of General Medical Sciences (R35GM119582) and the Defense Advanced Research Projects Agency.
The content is solely the responsibility of the authors and does not necessarily represent the official views of NIGMS, the National Institutes of Health, or the Defense Advanced Research Projects Agency.

\begin{supplement}

The data and code used to train equal, static, and adaptive ensembles can be found at \url{https://github.com/tomcm39/adaptively_stacking_ensembles_for_infleunza_forecasting_with_incomplete_data}.
Influenza-like illness, component model forecasts, and ensemble performance metrics are published on Harvard's Dataverse.
Links to the data are provided on the above GitHub page.
\end{supplement}

\begin{appendix}

\section{Computing $q(\pi)$ and $q(Z)$ for the Degenerate Variational Mixture model (deVI-MM)} \label{sec:appendixA}

The functional forms for $q(Z)$ and $q(\pi)$ are computed.
Readers interested in more details, and theoretical background, should consult~\cite{rustagi1976variational,blei2017variational,bishop2006pattern,murphy2012machine}.
In particular, \cite{blei2017variational} gives a brief introduction to variational inference focused on the applied statistician, while \cite{bishop2006pattern,murphy2012machine} provide more theoretical details. 
The most detailed material can be found in \cite{rustagi1976variational}.

We find the $q(Z)$ and $q(\pi)$ that maximize the lower bound $\mathcal{L}(q)$ by taking advantage of our factored $q$ and using iterated expectations.
\begin{align}
  \mathcal{L}(q) = E_{\pi,Z} \l\{ \log\l[p(\mathcal{D},Z,\pi)\r] - \log[q(\pi)] - \log[q(Z)]\r\}
\end{align}
Maximizing $q(\pi$), we can take the iterated expectation
\begin{align*}
  \max_{q(\pi)} \mathcal{L}(q) =  &\max_{q(\pi)} E_{\pi|Z} E_{Z} \l\{ \log\l[p(\mathcal{D},Z,\pi)\r] - \log[q(\pi)] - \log[q(Z)]\r\}\\
  &=\max_{q(\pi)} E_{\pi|Z}  \l\{ E_{Z}\log\l[p(\mathcal{D},Z,\pi)\r] - \log[q(\pi)]\r\}\\
  &=\min_{q(\pi)} \text{KL}\l\{ \log q(\pi) ||  E_{Z}\log\l[p(\mathcal{D},Z,\pi)\r] \r\}\\
\end{align*}
The Kullback-Leibler divergence, taking values from 0 to $\infty$, is minimized when
\begin{equation*}
    \log q(\pi) = E_{Z}\log\l[p(\mathcal{D},Z,\pi)\r]
\end{equation*}
or when 
\begin{equation*}
   q(\pi) \propto \exp\l\{E_{Z}\log\l[p(\mathcal{D},Z,\pi)\r]\r\}.   
\end{equation*}

Maximizing $Z$ follows a similar pattern.
The optimal hidden distributions of $q(\pi)$ and $q(Z)$ can then be computed
\begin{align*}
  q(\pi) &\propto \exp\l\{E_{Z}\log\l[p(D,Z,\pi)\r]\r\}\\
  q(Z) &\propto \exp\l\{E_{\pi}\log\l[p(D,Z,\pi)\r]\r\}.
\end{align*}

We first compute $q(\pi)$, expanding the complete loglikelihood, taking the expectation over $Z$, and recognizing this expectation as a specific distribution.
Here we don't explicitly describe $\pi$'s dependence on $t$ for convenience.
\begin{align*}
  \log q(\pi) &\propto \mathbbm{E}_{Z} \log \l[ p(D,Z,\pi)\r]\\
              &= \sum_{t=1}^{T} \sum_{m=1}^{M}  \mathbbm{E} \l[z(m,t)\r] \log \l[ \pi_m f_{m}(i_{t}) \r] + \sum_{m=1}^{M} \l[\alpha(t)-1\r] \log(\pi_{m}) - \log \l\{\eta\l[\alpha(t)\r]\r\}\\
              &= \sum_{t=1}^{T} \sum_{m=1}^{M}  \mathbbm{E} \l[z(m,t)\r] \log \l( \pi_m \r) + \sum_{m=1}^{M} \l[\alpha(t)-1\r] \log\l(\pi_m\r)\\
              &= \sum_{m=1}^{M} \log \l( \pi_m \r)  \l\{ \alpha(t) + \sum_{t=1}^{T}  \mathbbm{E} \l[z(m,t)\r] - 1 \r\}\\
              &= \sum_{m=1}^{M} \log \l( \pi_m \r)  \l\{ \alpha(t) + \sum_{t=1}^{T}  r(m,t) - 1 \r\},
\end{align*}
where $\eta$ is the normalizing constant for the Dirichlet distribution and $r(m,t)$ the expected value of the indicator variable $z_{m,t}$, the probability model $m$ generated the ILI value at time $t$.
Studying the form of $\log q(\pi), $we recognize $\pi$ is Dirichlet distributed
\begin{align*}
  q(\pi) &\sim \mathrm{Dir}\l( \gamma \r)\\
  \gamma\l[m\r] &= \alpha(t) + \sum_{t=1}^{T}  r(m,t)
\end{align*}

The same procedure can also be applied to compute $q(Z)$:
\begin{align}
  \log q(Z) &\propto \mathbbm{E}_{\pi} \log \l[ p(D,Z,\pi)\r] \nonumber \\
            &=  \sum_{t=1}^{T} \sum_{m=1}^{M}  z(m,t) \l\{\mathbbm{E}_{\pi} \log \l( \pi_m \r) + \log \l[ f_{m}(i_{t}) \r]\r\} +  \mathbbm{E}_{\pi}\sum_{m=1}^{M} \l[\alpha(t)-1\r] \log(\pi_{m}) - \log\l[\eta(\alpha)\r] \nonumber \\
            & \propto \sum_{t=1}^{T} \sum_{m=1}^{M}  z(m,t) \l\{ \mathbbm{E}_{\pi} \log \l[ \pi_m \r] + \log \l[ f_{m}(i_{t}) \r] \r\} \nonumber \\
  q(Z_{t,m}) &\propto  \exp \l \{ \mathbbm{E}_{\pi} \log \l( \pi_m \r) + \log \l[ f_{m}(i_{t}) \r] \r \}  \label{notNorm},
\end{align}
and we recognize $q(Z_{t,m})$ is Bernoulli distributed, and with the additional constraint that all indicators must sum to one for every time period $(t)$, we see $Z_{t}$ is multinomial for every $t$. 
\begin{align}
  q(Z_{t,m}) &\sim \mathrm{Bern}\l[ r(m,t) \r]\\
  r(m,t) &= \dfrac{ \exp \l \{ \mathbbm{E}_{\pi} \log \l( \pi_m \r) + \log \l[ f_{m}(i_{t}) \r] \r \} } {\sum_{m=1}^{M} \exp \l \{ \mathbbm{E}_{\pi} \log \l( \pi_m \r) + \log \l[ f_{m}(i_{t}) \r] \r \}}.
\end{align}

Although burdensome upfront, this approximate procedure drastically reduces computational time, compared to more intense monte carlo sampling techniques, and gives a good approximate answer, only assuming $Z$ and $\pi$ independent from one another.
Also note this mixture model algorithm (both EM and VI), unlike a typical Gaussian mixture model, cannot manipulate the paramters control the component model distributions.
But this inability to access the component model parameters opens up greater opportunities for other forecasters to submit models, requiring every forecast model to supply just 1, 2 ,3 , and 4 week ahead forecasts.

\section{Computing $E[\log(\pi)]$}
\label{app.expec}

Variational inference over hidden variables $(Z,\pi)$ requires computing the expected value of the log of $\pi$'s, a natural consequence of adapting the Dirichlet distribution to exponential form.
Exponential form rewrites a probability distribution in the form
\begin{equation}
  p\l[\pi|\eta(\alpha)\r] = h(\pi)e^{ \sum_{k} \eta_{k}(\alpha) T_{k}(\pi) - c(\eta)}.
\end{equation}
where $c(\pi)$ is a normalizing constant.
Two facts about exponential form lead to an analytic formula for $E(\log \pi)$: taking the gradient of $c(\pi)$ and relating the set of sufficient statistics $T(\pi)$ with this expectation.
Starting with the first fact.
If $c(\pi)$ is a normalizing constant, then it must equal
\begin{equation}
  c(\eta) = \log \l( \int  h(\pi)e^{ \eta^{'}(\alpha)T(\pi)} \; d \pi  \r),
\end{equation}
and it's gradient must equal
\begin{align}
   \nabla_{\eta} \l[c(\eta)\r] &= \int T_{k}(\pi) \f{h(\pi)e^{ \eta^{'}(\alpha)T(\pi)}}{\int  h(\pi)e^{ \eta^{'}(\alpha)T(\pi) }} \; d \pi\\
                &= \int T(\pi) p(\pi) = E[T(\pi)] .
\end{align}
A powerful consequence of exponential form, the gradient of the normalizing constant equals the expected value of the distribution's sufficient statistics.
The Second fact.
Working with the log likelihood of a distribution in exponential form, 
\begin{equation}
  \log \l[ \prod_{n=1}^{N} p(\pi|\alpha)\r] = N \log h(\pi) +  N \sum_{k} \eta_{k}(\alpha_{k})T_{k}(\pi) - N c(\pi),
\end{equation}
taking the gradient and setting equal to 0,
\begin{align}
  \nabla_{\eta} \log \l[ \prod_{n=1}^{N} p(\pi|\alpha)\r] &=  \sum T(\pi_{n}) - N \nabla c(\eta) = 0\\
  \nabla c(\eta) &= E\l[ T(\eta) \r] = \f{1}{N} T(\pi).
\end{align}
The expected value of the distribution's sufficient statistics are equal to the gradient of $c(\eta)$.
If the Dirichlet's sufficient statistics take the form $\log(\pi)$, then we only need to take the gradient of the normalizing constant to find an analytic expression.

Looking at the Dirichlet distribution, the loglikelihood equals
\begin{equation}
  \log \l [ \prod p(\pi_{n}|\alpha) \r] = N \log \Gamma \l(\sum_{\alpha} \alpha_{k}\r) - N \sum_{k} \log \Gamma \l(\alpha_{k}\r) + N \sum_{k} (\alpha_{k}-1) \log(\pi_{k})
\end{equation}
the sufficient statistics for $\pi_{k}$ take the form $\log \l( \pi_{k}\r)$.
Taking the gradient then will provide an analytic expression for computing the $\log \pi$'s expected value.
\begin{equation}
 \nabla_{\eta} \log \l [ \prod p(\pi_{n}|\alpha) \r] = N \psi \l(\sum_{\alpha} \alpha_{k}\r) - N \sum_{k} \psi \l(\alpha_{k}\r) + N \log(\pi_{k}),
\end{equation}
where $\psi$ is the digamma function.
Then the expected value of $\log\l(\pi\r)$ equals a difference of digamma functions
\begin{equation}
  E\l[\pi_{k}\r] = \psi\l(\alpha_{k}\r) -  \psi \l(\sum_{\alpha} \alpha_{k}\r).
\end{equation}
This formula can be used to compute the responsibility function (see Algorithm $2$ step $7$).

\section{Convexity analysis}
\label{convexity}

The problem of finding an optimal set of mixture weights can be recast as a constrained optimization problem.
Attempting to optimize a convex function over a convex set will prove a global optima exists, and showing strict convexity will prove the a unique vector obtains this optimum.

The original problem searches for weights that maximize a loglikelihood whose sum is constrained to equal one,
\begin{align}
    &\text{max} \sum_{t=1}^{T} \log \l[ \sum_{m=1}^{M} \pi_mf_{m}(i_{t}) \r] \label{oFunc}\\
    &\text{subject to} \nonumber \\ 
    &   \sum_{m=1}^{M} \pi_{m} = 1 \nonumber,
\end{align}
and can be converted to a Langragian with a single constraint $(\lambda)$
\begin{align}
  \mathcal{L}(\pi,\lambda) = \sum_{t=1}^{T} \log \l[ \sum_{m=1}^{M} \pi_mf_{m}(i_{t}) \r]  +\lambda \l( \sum_{m=1}^{M} \pi_m-1 \r) \nonumber.
\end{align}

Knowing the solution $\pi$ needs to lie in the $M$ dimensional simplex, a convex set, we only need to show the above function~\eqref{oFunc} is convex.
Given~~\eqref{oFunc} is differentiable at least twice, we will appeal to a second-order condition for convexity---proving the Hessian is positive semidefinite.
After proving the Hessian is positive semidefinite, going further and showing the Hessian is in fact positive definite will prove a unique vector $\pi$ is a the global optimum of our objective function~\eqref{oFunc}.

First, we compute the $m^{\text{th}}$ element of the gradient for \eqref{oFunc},
\begin{align*}
  \nabla_{\pi_m} f(\pi) = \sum_{t=1}^{T} \f{f_{m}(i_{i})}{\sum_{m=1}^{M} \pi_mf_{m}(i_{t}) }.
\end{align*}
Then the $(m,n)$ entry of the Hessian is
\begin{align*}
  H(m,n) = - \sum_{t=1}^{T} \f{ f_{m}(i_{t}) f_{n}(i_{t})}{ \l[\sum_{m=1}^{M} \pi_mf_{m}(i_{t})  \r]^{2} }.
\end{align*}
The Hessian is always negative, and if we minimized the negative loglikelihood (instead of maximizing the positive logelikihood) would see the Hessian is a positive semidefinite matrix.
But we can prove more by rewriting $H$
\begin{align*}
 H = F'F,
\end{align*}
where the $[m,t]$ element
\begin{equation*}
  F[m,t] = \f{f_{m}(i_{t})}{\sum_{m=1}^{M}\pi_mf_{m}(i_{t})}.
\end{equation*}
Positive definite matrices can always be written as a transposed matrix times the matrix itself, and so $H$ must also be positive definite.
Our convex optimization problem must then have a global optimum, attained by a unique vector $\pi$.
The inability to change forecasting models is a limitation, but allows us to guarantee a global optima.
When the component model parameters can also be updated, no such guarantee exists.

\clearpage
\section{Simplex plots of all Seasons}
\label{simplexplot}

Regularizing the adaptive ensemble suppresses large swings in weight assignment, independent of season.
The \adaptOver ensemble stays closer to equal weighting throughout the season compared to the \adaptOpt and \adaptNon ensembles.
If one exists, there is no strong relationship between the adaptive and static ensemble weighting.
Some seasons~(2014/2015 and 2016/2017) show higher variability in weight assignment than others.
The optimal weight assignments given data from all previous seasons (Static MLE) does not carry forward to the optimal weight assignment for the adaptive ensemble at the end of the current season.  

\begin{figure}[ht!]
    \centering
    \includegraphics[scale=0.90]{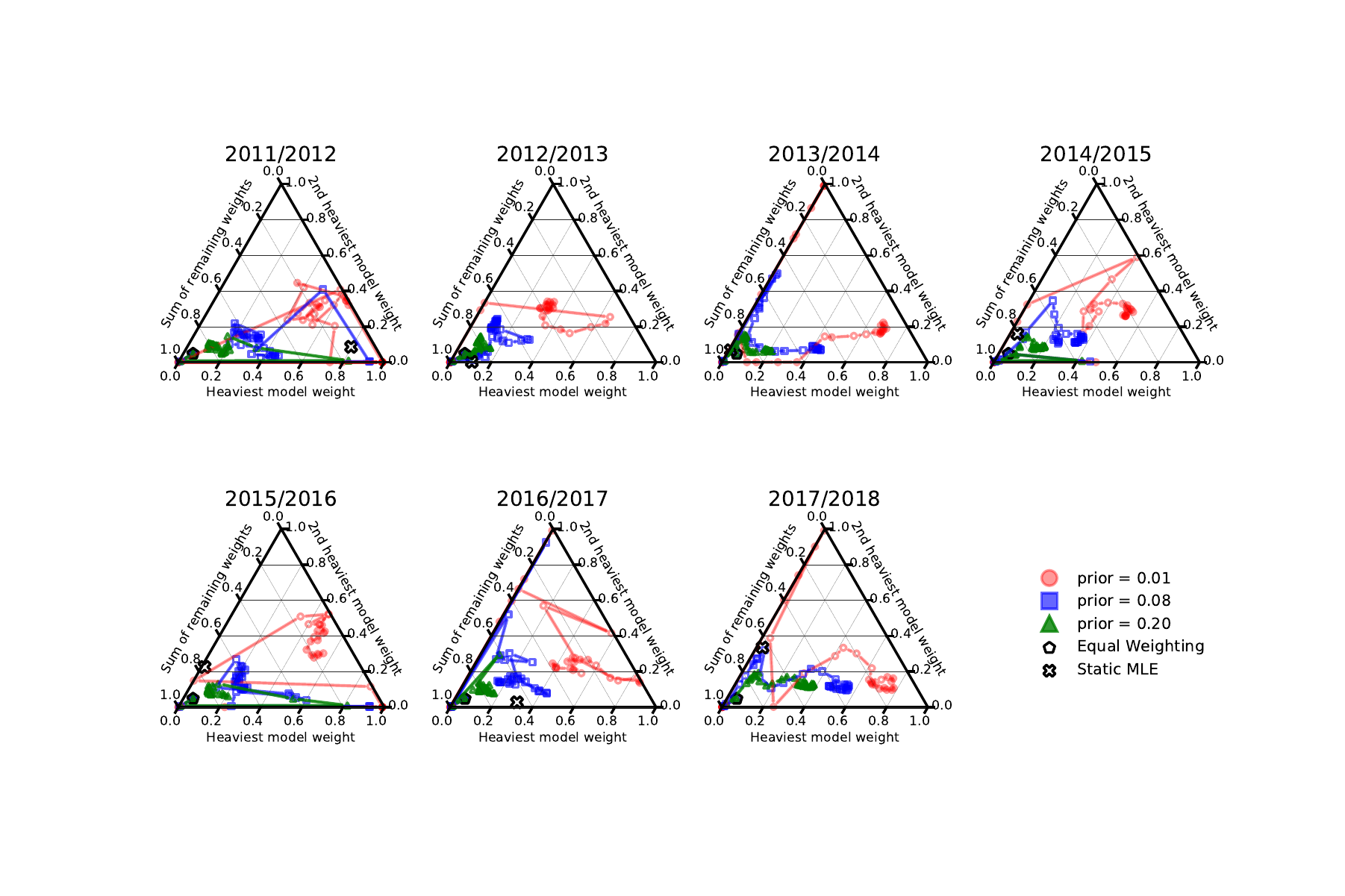}
    \caption{Component model weights for an adaptive ensemble with 1\% (non), 8\% (opt), and 20\% (over) priors plotted over epidemic week and stratified by season.
    Equal weighting is represented by a pentagon and the Static ensemble weights represented by an X.
    All ensembles start at an equal weighted triple of the 1st and 2nd highest weighted component model, and the sum of the remaining components at the end the season ($\pi_{t}^{(1)}$,$\pi_{t}^{2}$,$\sum_{j=3}^{M}\pi_{t}^{(j)}$), and move from week to week as new data is recieved and the adaptive ensemble re-estimates component model weight assignments.
        \label{fig.bagOfDoritos}}
\end{figure}

\clearpage
\section{Tabulated comparisons of adaptive, static, and equal logscores}
\label{tbRegress}

The \adaptOpt ensemble consistently outperforms the EW ensemble, and performs similarly to the static ensemble.
When compared to the EW model, the adaptive cannot show statistical significance in the 2012/2013 season, but all differences are in the positive direction, favoring the adaptive model.
The adaptive and static comparisons are more even.
Some season, regions, and targets favor the static ensemble, others the adaptive model, but absolute differences are small.
Despite markedly less data, the adaptive ensemble bests the EW ensemble and shows similar performance to the static.

\begin{table}[ht!]
  \begin{tabular}{lcrrcrr}
    \hline
     & \multicolumn{3}{c}{\textbf{Adaptive$_{\text{opt}}$ - EW}} & \multicolumn{3}{c}{\textbf{Adaptive$_{\text{opt}}$ - Static}}\\
     & $\beta$ (95CI) & $p$ & $p_{\text{permutation}}$ & $\beta$ (95CI) & $p$ & $p_{\text{permutation}}$\\
     \hline
     
  Season\\
 \hspace{3mm} 2011/2012 & 0.13 (0.08, 0.19) & $<$0.01  &$<0.01$& 0.02 (-0.04, 0.08) &0.51&0.58\\
 \hspace{3mm} 2012/2013	& 0.06 (0.00, 0.12) &0.04&0.21& 0.02 (-0.04, 0.08) &0.48&0.52\\
 \hspace{3mm} 2013/2014	& 0.10 (0.04, 0.15) & $<$0.01  &0.04& 0.00 (-0.06, 0.06) &0.96&0.97\\
 \hspace{3mm} 2014/2015	& 0.14 (0.09, 0.20) & $<$0.01  &$<0.01$& 0.03 (-0.03, 0.09) &0.30&0.36\\
 \hspace{3mm} 2015/2016	& 0.13 (0.07, 0.18) & $<$0.01  &0.01& -0.02 (-0.08, 0.04) &0.54&1.00\\
 \hspace{3mm} 2016/2017	& 0.11 (0.06, 0.17) & $<$0.01  &0.01& -0.04 (-0.10, 0.02) &0.20&1.00\\
 \hspace{3mm} 2017/2018	& 0.21 (0.15, 0.26) & $<$0.01  &0.00& -0.01 (-0.06, 0.05) &0.86&1.00\\ 

 Region\\    
 \hspace{3mm}	HHS1 &  0.16 (0.11, 0.22) & $<$0.01  &$<0.01$& 0.01 (-0.05, 0.07) &0.80 &0.83\\
 \hspace{3mm}	HHS2 &  0.11 (0.06, 0.17) & $<$0.01  &$<0.01$& -0.01 (-0.07, 0.06) &0.85&1.00\\
 \hspace{3mm}	HHS3 &  0.13 (0.08, 0.18) & $<$0.01  &$<0.01$& 0.04 (-0.03, 0.10) &0.28&0.41\\
 \hspace{3mm}	HHS4 &  0.12 (0.07, 0.17) & $<$0.01  &$<0.01$& -0.01 (-0.08, 0.05) &0.68&1.00\\
 \hspace{3mm}	HHS5 &  0.13 (0.07, 0.18) & $<$0.01  &$<0.01$& -0.05 (-0.11, 0.01) &0.12&1.00\\
 \hspace{3mm}	HHS6 &  0.12 (0.07, 0.18) & $<$0.01  &0.001& -0.04 (-0.11, 0.02) &0.17&1.00\\
 \hspace{3mm}	HHS7 &  0.13 (0.08, 0.18) & $<$0.01  &$<0.01$& 0.05 (-0.01, 0.12) &0.09&0.20\\
 \hspace{3mm}	HHS8 &  0.14 (0.09, 0.20) & $<$0.01  &$<0.01$& 0.00 (-0.06, 0.06) &0.99&1.00\\
 \hspace{3mm}	HHS9 &  0.08 (0.02, 0.13) & $<$0.01  &0.01& 0.02 (-0.04, 0.08) &0.55&0.60\\
 \hspace{3mm}	HHS10 & 0.12 (0.07, 0.17) & $<$0.01  &$<0.01$& -0.04 (-0.11, 0.02) &0.19&1.00\\
 \hspace{3mm}	Nat	 &  0.13 (0.07, 0.18) & $<$0.01  &$<0.01$& 0.06 (-0.01, 0.12) &0.08&0.17\\

 Target\\
 \hspace{3mm}	1 week ahead	&   0.16 (0.11, 0.22) & $<$0.01  &$<0.01$& 0.06 (0.00, 0.11) &0.07&0.18\\
 \hspace{3mm}	2 week ahead	&  0.13 (0.08, 0.19) & $<$0.01  &$<0.01$& -0.01 (-0.07, 0.05) &0.81&1.00\\
 \hspace{3mm}	3 week ahead	&  0.11 (0.06, 0.17) & $<$0.01  &$<0.01$& -0.01 (-0.07, 0.05) &0.67&1.00\\
 \hspace{3mm}	4 week ahead	&  0.09 (0.03, 0.14) & $<$0.01  &0.02& -0.03 (-0.09, 0.03) &0.35&1.00\\
 \hline
\end{tabular}

\caption{
Random effects regressions compared logscores between the \adaptOpt vs equally-weighted and \adaptOpt vs static ensembles.
The model included an intercept, and separate random effect for: season, region, and target.
The dependent variables is the difference in logscores paired by season-region-target-epidemic week.
Conditional mean, 95\%CI, asymptotic, and a permutation based p-value are reported.}

\end{table}







\clearpage
\section{Regularization improves static ensemble}

\begin{figure}[ht!]
    \centering
    \includegraphics[scale=0.60]{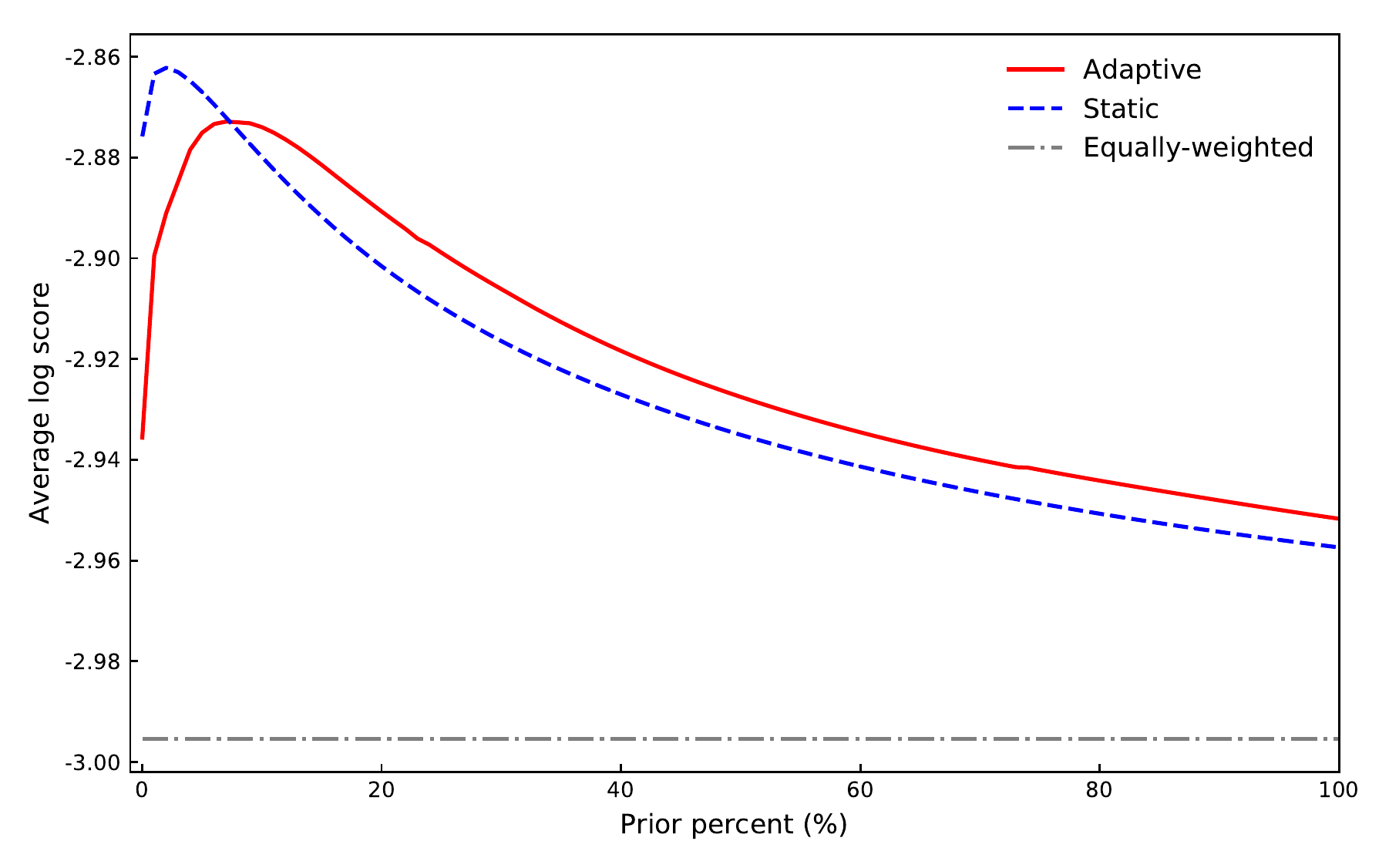}
    \caption{The average logscore for the equally-weighted ensemble, and adaptive, static ensembles for prior from 0\% to 100\% by 1\%.
    The logscore is averaged over season, region, and target.
    The highest average logscore corresponds to a larger prior than the static ensemble. But both the static and adaptive ensemble benefit from regularization.\label{fig.staticHelp}}
\end{figure}

Regularizing ensemble weights improves both adaptive and static ensemble performance~(Fig.~\ref{fig.staticHelp}). 
The static ensemble achieves peak performance with a smaller prior than the adaptive.
This smaller prior reflects the data the static model is trained on. 
Past seasons of finalized ILI data. 
The adaptive model, finding peak performance for a larger prior, needs to account for the lack of training data and the revision-prone state of the data mid-season.

\clearpage
\section{Supplementary Figure}
\begin{figure}[ht!]
    \centering
    \includegraphics[scale=0.60]{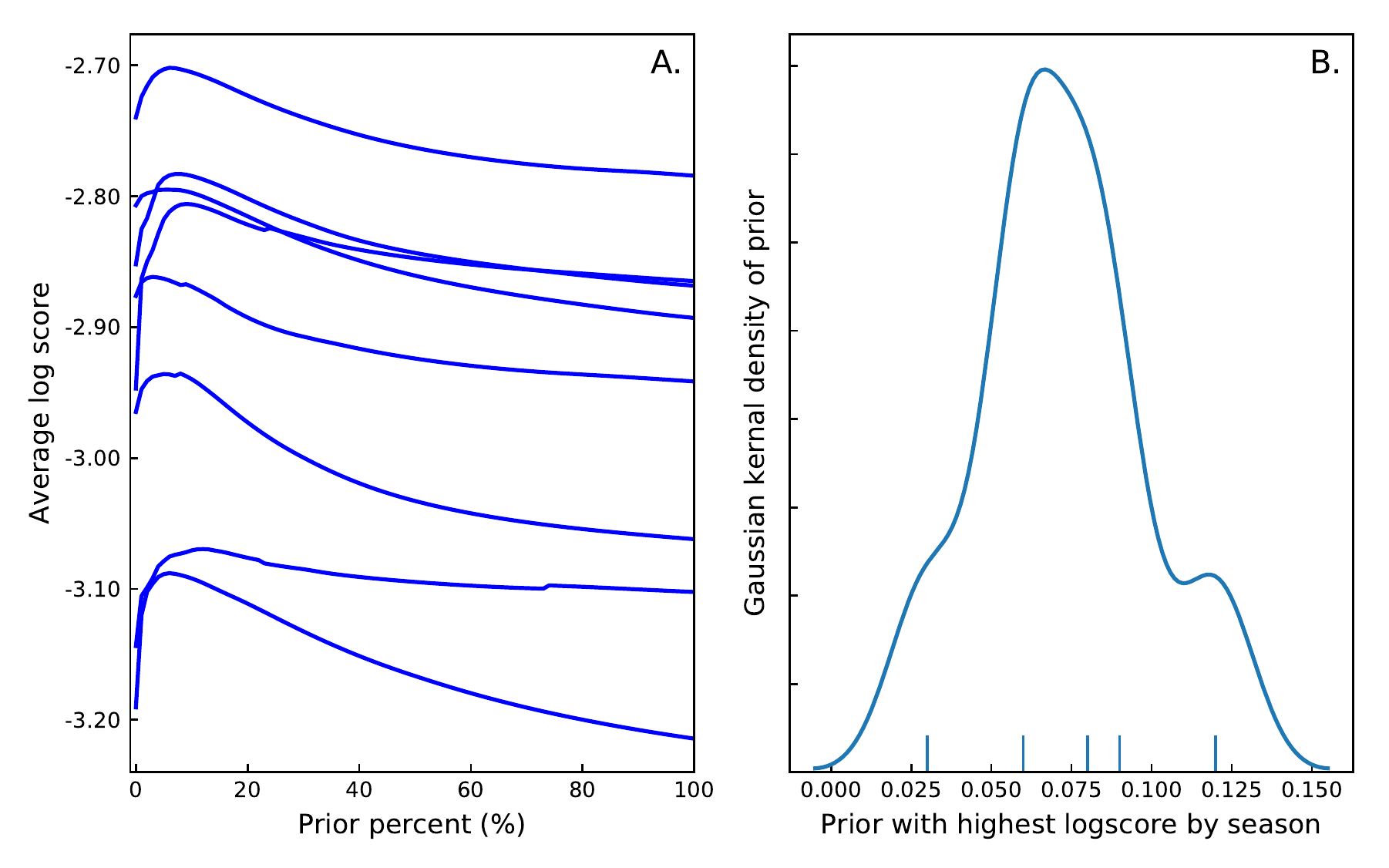}
    \caption{
      (A.)~The adaptive ensemble is fit for prior percentages from 0\% to 100\% by 1\% and the average logscore is computed, and stratified by seasons 2010/2011-2017/2018.
      (B.)~ the distribution of priors corresponding to the highest logscore per season.
      (A.)~Each seasons shows a similar trend in adaptive ensemble logscore using different prior percentages.
      Priors close to 0\% produce small logscores, a peak logscore occurs near a prior of 8\%, and then logscores decrease with larger priors.
      (B.)~The 25th and 75th percentile of priors corresponding to peak logscores are 6\% and 8.25\%.
      The smallest prior equals 3\% and largest equals 12\%.
      The large probability near a prior percentage of 8\%, across influenza seasons, suggests an optimal weighting may lie near 8\%, was not specific to the 2010/2011 season analysis, and could work well as a prior for future seasons.
      \label{suppl1.logScoresPerSeason}}
  \end{figure}

\end{appendix}

\clearpage
\bibliographystyle{imsart-number}


\end{document}